\documentclass[12pt]{article}

\usepackage{amsfonts,amssymb,amsmath,amsthm}
\usepackage[dvips, final]{graphics}
\usepackage{epsfig}
\usepackage{pinlabel}
\usepackage{url}
\usepackage{verbatim}

\theoremstyle{plain}
\newtheorem{Theorem}{Theorem}[section]
\newtheorem{Definition}[Theorem]{Definition}
\newtheorem{Lemma}[Theorem]{Lemma}
\newtheorem{Remark}[Theorem]{Remark}
\newtheorem{Proposition}[Theorem]{Proposition}

\newtheorem{Example}[Theorem]{Example}

\theoremstyle{definition}
\newtheorem*{Ack}{Acknowledgment}

\newcommand{\R}{\mathbb R}

\newcommand{\C}{\mathbb C}
\newcommand{\J}{\operatorname J}
\newcommand{\Ji}{\operatorname J^{-1}}

\newcommand{\Mic}{{\bf MiC}}
\newcommand{\Pois}{{\bf Poiss}}
\newcommand{\ExtSym}{{\bf ExtSympl}}

\newcommand{\Mon}[1]{{\bf Mon}(#1)}
\newcommand{\Deq}{{\bf D}}

\newcommand{\Sympl}{\bf {Sympl}}

\newcommand{\Cot}[1]{\mathrm{T}^*{#1}}
\newcommand{\Tan}[1]{\mathrm{T}{#1}}
\newcommand{\Ver}[1]{\mathrm{V}{#1}}

\newcommand{\id}{\operatorname{id}}
\newcommand{\morph}{\operatorname{Hom}}

\newcommand{\graph}{\operatorname{graph}}
\newcommand{\stat}[1]{{\bf Stat}_{(#1)}}

\newcommand{\neutral}{\textbf{E}}

\newcommand{\neutralmorph}{\textbf{e}}

\newcommand{\cH}{\mathcal{H}}

\begin{document}

\vspace{3em}
\begin{center}

{\Large{\bf Cotangent Microbundle Category, I }}

\vspace{3em}

\textbf{ Alberto S. Cattaneo$^{1}$, Benoit Dherin$^{2}$  and Alan Weinstein$^{2}$}

\vspace{1em}

${}^1$Institut f\"ur Mathematik, Universit\"at Z\"urich--Irchel\\
Winterthurerstrasse 190, CH-8057 Z\"urich, Switzerland \\[1em]

${}^2$Department of Mathematics, University of California\\
Berkeley, California 94720-3840\\[1em]

\end{center}

\vspace{1em}

\begin{abstract}
We define a local version of the extended symplectic
category, the  cotangent microbundle category,
$\Mic$, which turns out to be a true monoidal category. 
We show that a monoid
in this category induces a Poisson manifold together with
the local symplectic groupoid 
integrating it. Moreover, we prove that 
monoid morphisms produce Poisson maps between the induced Poisson manifolds
in a functorial way. This gives a functor between the category
of monoids in $\Mic$ and the category of Poisson manifolds
and Poisson maps. Conversely, the semi-classical part of the
Kontsevich star-product associated to a real-analytic Poisson structure on an open
subset of $\R^n$ produces a monoid in $\Mic$.  
\end{abstract}

\tableofcontents

\section{Introduction}

There is a category $\Sympl$ whose objects are
finite-dimensional symplectic manifolds $(M,\omega)$
and whose morphisms are symplectomorphisms
$\Psi:(M,\omega_M)\rightarrow (N,\omega_N)$. 
In attempting to understand the quantization
procedure of physicists from a mathematical perspective, one
may think of it as a functor from this symplectic category,
where classical mechanics takes place,
into the category of Hilbert spaces and unitary operators,
which is the realm of
quantum mechanics.  It is well known that this symplectic category is
too large, since there are ``no-go'' theorems which show that the
group of all symplectomorphisms on $(M,\omega)$ does not act in a
physically meaningful way on a corresponding Hilbert space.  One
standard remedy for this is to replace $\Sympl$ by a smaller category,
replacing the symplectomorphism groups by certain finite-dimensional
subgroups.  Another is to replace the Hilbert spaces and operators by
objects depending on a formal parameter.

But there is also a sense in which the category $\Sympl$ is too {\em small},
since it does not contain morphisms corresponding to operators such as
projections and the self-adjoint
(or skew-adjoint) operators which play the role of observables in
quantum mechanics, nor can it encode the algebra structure itself on
the space of observables.   (This collection of observables is not actually
a Hilbert space, but certain spaces of operators do carry a
vector space structure, with the
inner product associated to the Hilbert-Schmidt norm.)

To enlarge the symplectic category, we look to the 
``dictionary'' of quantization, following, for example,
\cite{weinstein1996}.
In this dictionary,
the cartesian product of
symplectic manifolds corresponds to the tensor product of Hilbert
spaces, and replacing a symplectic manifold $(M,\omega)$ by
$(M,-\omega)$ (which we denote by $\overline M$ when we omit the
symplectic structure from the notation for a given symplectic manifold)
corresponds to replacing a Hilbert space $\cH$ by its conjugate, or dual,
space $\cH ^*$.  Thus, if symplectic manifolds $M_1$ and $M_2$
correspond to Hilbert spaces $\cH_1$ and $\cH_2$
the product $\overline{M}\times N$ 
corresponds to $\cH_1 ^* \otimes\cH_2$, which, with a suitable definition
of the tensor product, is the space $L(\cH_1,\cH_2)$ of all linear
operators from $\cH_1$ to $\cH_2$.

Another entry in the dictionary says that lagrangian submanifolds
in symplectic manifolds (perhaps carrying half-densities) correspond
to vectors or lines in Hilbert space.  Combining this idea with that
in the paragraph above, we conclude that lagrangian submanifolds in
$\overline{M}\times N$ should correspond to linear operators from 
$\cH_1$ to $\cH_2$. 

This suggests that, if the space of observables $\cH$ for a quantum system
corresponds to a symplectic manifold $M$, then the algebra structure
on $\cH$ should be given by a lagrangian submanifold $\mu$ in $\overline{M}
\times \overline{M} \times M.$  The algebra axioms of unitality and
associativity should be encoded by monoidal properties of $\mu$ in an
{\bf extended symplectic category}, $\ExtSym$, where the morphisms from $M$
to $N$ are the {\bf canonical relations}, i.e. all the
lagrangian submanifolds of $\overline M\times N$ (not just
those which are the graphs of symplectomorphisms) and where
the morphism composition is the usual composition of relations
\footnote{In the context of symplectic geometry, the composition of
canonical relations
may be seen as a special instance of symplectic reduction.
Consider $C:= \overline M\times\Delta_N\times P$, where
$\Delta_N$ is the diagonal subset of $\overline N\times N$. $C$ is
a coisotropic submanifold of $\overline M\times N\times\overline N
\times P$ and $L_2\circ L_1$ happens to be the reduction of the 
lagrangian submanifold $L_1\times L_2$ with respect to $C$.
Thus, if $L_1\in \subset\overline M\times N$ and
$L_2\subset \overline N\times P$ are lagrangian submanifolds, then
$L_2\circ L_1$ is a lagrangian submanifold of $\overline M\times P$
whenever it is a submanifold.}.
However, a problem immediately occurs: the composition of canonical
relations may yield relations which are not submanifolds anymore and thus,
not canonical relations!  $\ExtSym$ is then not a true category, as
the morphisms can not be always composed.
It is thus rather uncomfortable to speak about a quantization functor
in this context. 

There have already been several approaches to remedy this defect.
One, by Guillemin and Sternberg \cite{GS1979}, is to consider
only symplectic vector spaces and linear canonical relations.
Another, by Wehrheim and Woodward \cite{WW2007}, is to enlarge the category still
further by allowing arbitrary ``formal'' products of canonical
relations, and equating them to actual products when the latter exist
as manifolds.

In this paper, we take another approach.
We define a local version of the extended
symplectic category which is
a true category.  We restrict ourselves to
cotangent bundles with their canonical symplectic structures
and define
$\morph(\Cot M,\Cot N)$ to be germs near the zero section of
canonical relations which are suitably close to the conormal
bundles of graphs of diffeomorphisms from $N$ to $M$. We call
the resulting category the {\bf cotangent microbundle category}.
We choose this name for the category since the objects involved
are symplectic version of the microbundles
introduced by Milnor in \cite{milnor1964}.

In Section \ref{sec:trans}, we
express, in terms of  transversality,
the condition that germs of lagrangian submanifolds
are somehow close to the conormal bundle of the graph
of a map between the bases.

In Section \ref{sec:cat}, we define the cotangent microbundle category
$\Mic$, by allowing the morphisms $\morph(\Cot M,\Cot N)$
to be the transverse lagrangian germs in $\overline{\Cot M}\times \Cot N$
as defined in Section \ref{sec:trans}. We show that the composition 
is always well-defined and that the resulting category
is a true monoidal category. Let us note here that the lagrangian
operads considered in \cite{Cattaneo2002} 
and in \cite{CDF2005bis} are closely related
to the endomorphism operad associated to any object in $\Mic$ 
in the usual way. This will be the subject of  future work.

In Section \ref{sec:gen}, we describe each morphism
locally in terms of a single function: the generating
function of the transverse lagrangian germ. We derive
a composition formula for generating functions and
show how they behave under changes of charts.

In Section \ref{sec:gpd}, we prove that a 
monoid $(\Cot M,\mu,\neutralmorph)$ in 
the cotangent microbundle category induces a Poisson 
structure on the base $M$ together with a
local symplectic groupoid $(s,t):\Cot M\rightrightarrows M$
integrating it. All the induced structures are described
explicitly in terms of generating functions. We show
that isomorphisms of monoids produce
Poisson diffeomorphisms between the induced Poisson
structures and local groupoid isomorphisms between
the induced local symplectic groupoids. 
This gives a functor from the category of 
monoids in $\Mic$ to the category $\Pois$ of Poisson
manifolds. These results are very much in the line
of the ``categories" introduced by Zakrzewski in \cite{zakrzewski1990}
and studied by Crainic and Fernandes in \cite{CF2006}. 

Section \ref{sec:ex} is devoted to explicit examples
of monoids in $\Mic$, their induced Poisson 
structures and local symplectic groupoids. In particular,
we give the generating function that induces
the symplectic Poisson structure, the generating
function that induces the Kirillov--Kostant Poisson
structure on the dual of a Lie algebra - the generating
function is the Baker--Campbell--Hausdorff formula in this
case - and the generating function attached to an analytical
Poisson structure on open subset of $\R^d$. The latest
generating function encompasses the two previous ones.
It is given by the semi-classical part of Kontsevich's
star-product. This last example supports the hope
that the cotangent microbundle category is the right
framework to construct a quantization functor. 

\begin{Ack}We thank Giovanni Felder for the many fruitful discussions
we had on the subject and Domenico Fiorenza and Jim Stasheff for having
read carefully a previous paper, ``The formal lagrangian operad'', which has
led to this one and for having shared their ideas. 
A.S.C. acknowledges partial support of SNF Grant 20-113439, of the European
Union through the FP6 Marie Curie RTN ENIGMA (contract number
MRTN-CT-2004-5652), and of the European Science Foundation through
the MISGAM program.  B. D.  acknowledges partial support of SNF Grant No.~PA002-113136 and
the mathematics departments of  Geneva University and UC Berkeley,
where part of this project was conducted. A.W. acknowledges partial
support from NSF grant No.~DMS-0707137.
\end{Ack}

\section{The transversality condition}\label{sec:trans}
The extended symplectic category is not a true category, as morphisms
can not always be composed. In order to obtain a true category,
we may restrict to special classes
of symplectic manifolds and special classes of lagrangian
submanifolds $L\subset \overline M\times N$ so that the composition is always
well-defined. Guillemin and Sternberg used linear symplectic spaces
and lagrangian linear subspaces, but this is too restrictive for most
purposes. In this article, we consider the cotangent bundle category
and modify it carefully. The objects in the cotangent 
bundle category are cotangent bundles $\Cot M$ over 
smooth manifolds $M$ endowed with their canonical symplectic
structure. Naively, a morphism $\Psi:\Cot M\rightarrow \Cot N$ is
a symplectomorphism that respects the zero sections.
For further generalization purposes, we will
rename the zero-section of a cotangent bundle $\Cot M$, the
{\bf core} of $\Cot M$ and we will refer to it as either $Z_M$ 
or simply as $M$.
Thus, a morphism in the cotangent bundle category preserves
the core.

In this section, we reformulate this property (preserving the
core) in
a way that it may be applied to
general lagrangian submanifolds $L$ of $\overline{\Cot M}\times \Cot N$
and not only to the graphs of symplectomorphisms. We call
this condition the {\bf transversality condition}.
As this condition is a local one (it concerns only
a neighborhood of $M\times N$ in $\Cot M\times \Cot N$
exactly as the condition $\Psi(M) = N$), we are
led to consider germs of lagrangian submanifolds.

\begin{Definition}
We say that a diffeomorphism 
$\Psi:\Cot M\rightarrow \Cot N$ covers
a map $\phi:N\rightarrow M$ if
$\Psi(0,\phi(x)) = (0,x)$ for all $x\in M$.
\end{Definition}

Note that this seems to say that $\Psi$
extends $\phi^{-1}$. Later, we will need 
to allow situations where $\phi$ 
not invertible.

\begin{Lemma}\label{lem:cotbd}
Let $\Psi:\Cot M\rightarrow \Cot N$ be a 
symplectomorphism preserving the cores, i.e., such that $\Psi(M) = N$.
Then:
\begin{itemize}
\item[(1)] There exists a unique map $\phi:N\rightarrow M$ such
that $\psi$ covers $\Phi$.
\item[(2)] In any local chart of the form $U=\Cot U_1\times
\Cot U_2$ of $\Cot M\times\Cot N$ , there exists a neighborhood 
$V$ of $U_1\times U_2$ in $U$
where the graph of $\Psi$ is of the form:
$$\graph\Psi\cap V = 
\Bigg\{
\bigg(\big(p_1,G(p_1,x_2)\big),
\big(H(p_1,x_2),x_2\big)\bigg): (p_1,x_2)\in W
\Bigg\},$$
where $p_1,x_1$ and $p_2,x_2$ are the local 
coordinates of $\Cot U_1$ and $\Cot U_2$ respectively,
$W$ is a neighborhood of $\{0\}\times U_2$ in
$(\R^d)^*\times U_2$ and $G:W\rightarrow U_1$,
and $H:W\rightarrow (\R^d)^*$ are smooth maps
such that $G(0,x_2) = \phi(x_2)$ and $H(0,x_2) = 0$.
\end{itemize}
\end{Lemma}

\begin{proof}
As $\Psi$ respects the core,
its restriction to $M$ induces a map
$g:=\Psi|_M$ from $M$ to $N$. Since $\Psi$
is a diffeomorphism, the induced map $g$ is
invertible. We denote by $\phi$ the inverse of $g$.
Clearly, $\phi$ is a diffeomorphism covered by $\Psi$.
%The graph of $\Psi$ is of the form
%jj$$
%\graph \Psi = \Bigg\{
%\big((p_1,x),(P,X)\big): \Psi(p,x) = (P,X)
%\Bigg\}.
%$$
In a local chart $U= \Cot U_1\times\Cot U_2$ of
$\Cot M\times\Cot N$,
let us write $\Psi$ as: $$\Psi(p_1,x_1) = \big(U(p_1,x_1),V(p_1,x_1)\big).$$
Then, for fixed $p_1$, consider the equation
\begin{eqnarray}\label{eq:xpart}
V(p_1,x_1) & = & x_2.
\end{eqnarray}
If $p_1=0$, then $V(0,x_1)= \phi^{-1}(x_1)$ and
$\nabla_{x_1}V(0,x_1) = \nabla\phi^{-1}(x_1)$
is invertible as $\phi$ is a diffeomorphism. The implicit
function Theorem tells us that we may invert equation
\eqref{eq:xpart}, i.e., we may find a function $G$ 
such that:
$$x_1 = G(p_1,x_2) \quad\textrm{s.t.}\quad
V(p_1,G(p_1,x_2)) = x_2,$$
for $(p_1,x_2)$ in a neighborhood $W$ of
$\{0\}\times U_2$ in $(\R^d)^*\times U_2$.
Thus, in a neighborhood $V$ of $U_1\times U_2$
in $\Cot U_1\times \Cot U_2$, we have that:
$$\graph \Psi\cap V =
\Bigg\{
\bigg(\big(p_1,G(p_1,x_2)\big),\big(H(p_1,x_2),x_2\big)\bigg):
\quad (p_1,x_2)\in W
\Bigg\},
$$
where $H(p_1,x_2) = U\big(p_1,G(p_1,x_2)\big)$.
Now, by definition, we have that,
$$ x_2 = V(0,G(0,x_2)) = \phi^{-1}(G(0,x_2)),$$
and thus $G(0,x_2) = \phi(x_2)$. On the other hand,
$$H(0,x_2) = U(0,G(0,x_2)) = 0.$$
\end{proof}

Let us express the content of Lemma \ref{lem:cotbd} in
a geometrical way.
Consider two cotangent bundles $\Cot M$ and $\Cot N$ and a map
$\phi$ from $N$ to $M$. Let $B_\phi$ be the pullback of
$\Cot M$ by $\phi$, i.e.,
$$B_\phi := 
\Bigg\{
(p_1,x_2):p_1\in \operatorname{T}_{\phi(x_2)}^* M,\quad x_2\in N
\Bigg\}
$$
and $Z_\phi$ its zero section.
Define $G_\phi:Z_\phi\rightarrow \overline{\Cot M}
\times \Cot N$ to be the map taking $Z_\phi$ to the
graph of $\phi$ in $M\times N$ considered as
a submanifold of $\overline{\Cot M}\times\Cot N$:
$$G_\phi(0,x) = \big((0,\phi(x)),(0,x)\big).$$ 

\begin{Definition}
A lagrangian embedding germ around $G_\phi$ is 
an equivalence class of lagrangian embeddings
$i_\phi:B_\phi\hookrightarrow \overline{\Cot M}\times\Cot N$
such that $i_\phi|_{Z_\phi} = G_\phi$, where two such lagrangian
embeddings are equivalent if there exists a neighborhood
$U$ of $Z_\phi$ in $B_\phi$ where their images coincide.
We denote the class of $i_\phi$ by $[i_\phi]$. When the context
is clear, we will use the $i_\phi$ to denote its class.
\end{Definition}

\begin{figure}[h!]
\labellist
\small\hair 2pt
\pinlabel ${\Cot M}$ at 0 10 
\pinlabel ${\Cot N}$ at 160 10
\pinlabel $M$ at 0 50 
\pinlabel $N$ at 160 50 
\pinlabel $x_2$ at 204 36 
\pinlabel $\phi(x_2)$ at 36 36
\pinlabel $\phi$ at 119 8
\pinlabel ${\Tan(Z_M)\times \Ver(\Cot N)}$ at 120 80
\endlabellist
\centering
\includegraphics[scale=1]{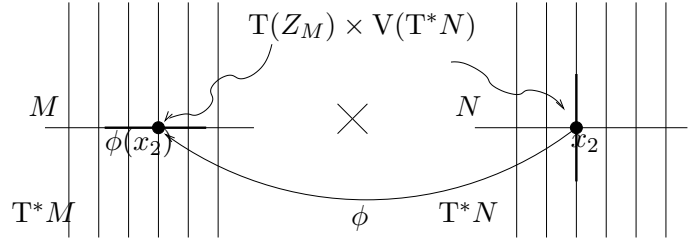}
\caption{The distribution $G_\phi^*\Lambda$ at the point $\Big((0,\phi(x_2)),(0,x_2)\Big)$
in $\Cot M\times \Cot N$.}
\label{fig:dist}
\end{figure}

The tangent bundle $\Tan(\Cot M\times \Cot N)$, restricted
to the product of the bases $M\times N$, has a natural
subbundle over $M\times N$:
$$\Lambda := \Tan (Z_M) \times \Ver(\Cot N),$$
where $\Tan (Z_M)$ is the tangent space to the
zero section $Z_M$ in $\Cot M$ and $\Ver(\Cot N)$ is
the tangent space to the vertical fibers in $\Cot N$.
We may pull back this bundle via the map $G_\phi$ to
a bundle $G_\phi^*\Lambda$ over $Z_\phi$, the zero
section of $B_\phi$. Figure \ref{fig:dist} represents a fiber of
this bundle $G_\phi^*\Lambda$ over a point in $\graph \phi =
G_\phi(Z_\phi)$.

\begin{Definition}\label{def:trans}
We call {\bf transverse lagrangian germ} a germ
$[i_\phi]$ of a lagrangian embedding $i_\phi:B_\phi\rightarrow
\overline{\Cot M}\times \Cot N$ around $G_\phi$
such that one (and thus any) of its representatives
$i_\phi$ is transverse to $G_\phi^*\Lambda$.
\end{Definition}

\begin{figure}[h!]
\labellist
\small\hair 2pt
\pinlabel $B_\phi$ at 7 38
\pinlabel $Z_\phi$ at 35 87
\pinlabel $i_\phi$ at 152 111
\pinlabel $j_\phi$ at 152 82
\pinlabel $G_\phi$ at 178 22
\pinlabel $\graph \phi$ at 317 20
\pinlabel $G_\phi^*\Lambda$ at 296 109
\endlabellist
\centering
\includegraphics[scale=1]{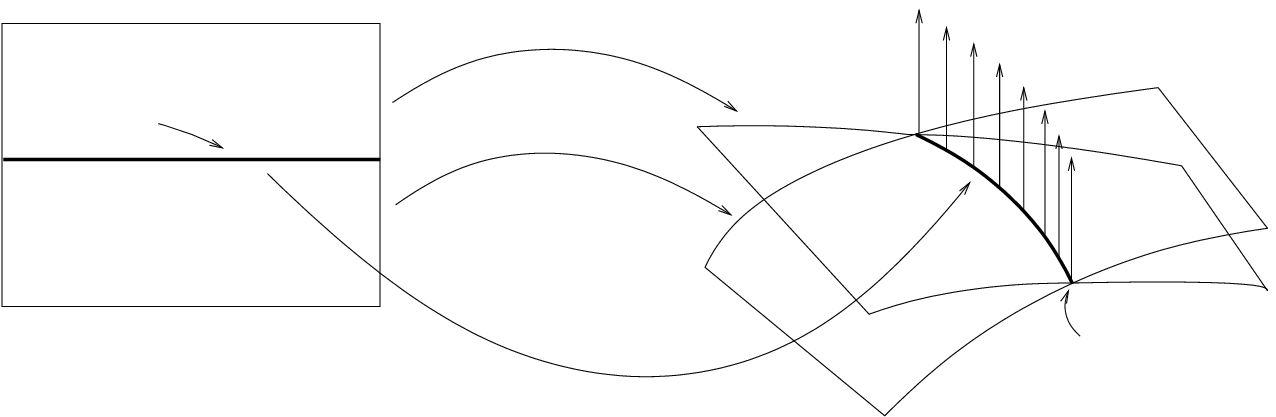}
\caption{Two transverse lagrangian germs $i_\phi$ and $j_\phi$ 
around  $G_\phi$.}
\label{fig:2trans}
\end{figure}

Figure \ref{fig:2trans} two represents such transverse germs around
the same core map $\phi$.

\vspace{2cm}

Let us see how this transversality condition translates
in local charts. 
Take $U_1$ a local chart of $M$ and $U_2$ a
local chart of $N$. Then $U=\Cot U_1\times\Cot U_2$
is a local chart of $\Cot M\times\Cot N$, and $B_\phi^U = \phi^*(\Cot U_1)$
is a local chart of $B_\phi$. Observe that these special local
charts cover a neighborhood of $M\times N$ in $\Cot M\times\Cot N$ and,
thus, are enough to describe completely germs of lagrangian embedding 
$i_\phi:B_\phi\hookrightarrow \overline{\Cot M}\times \Cot N$ around $G_\phi$.
Let us denote by $i_\phi^U:B_\phi^U\hookrightarrow U$ the representation
of $i_\phi$ in $U$. If the local coordinates on $U$ are $p_1,x_1,p_2,x_2$,
then the local coordinates on $B_\phi^U$ are $p_1,x_2$.

\begin{Lemma}\label{lem:trans}
A germ of lagrangian embedding $i_\phi:B_\phi\hookrightarrow
\overline{\Cot M}\times \Cot N$
is transverse to $G_\phi^*\Lambda$ iff, for any local chart $U$ as above, we
have that:
\begin{eqnarray}\label{belleforme}
i_\phi^U(W) & = & \Bigg\{
\bigg(\big(p_1,G(p_1,x_2)\big),
\big(H(p_1,x_2),x_2\big)\bigg): (p_1,x_2)\in W
\Bigg\},
\end{eqnarray}
where $W$ is a neighborhood of the zero section of $B_\phi^U$ and
$G:W\rightarrow U_1$ and $H:W\rightarrow (\R^k)^*$ are smooth maps such
that $G(0,x_2) = \phi(x_2)$ and $H(0,x_2) = 0$.
\end{Lemma}

\begin{figure}[h!]
\labellist
\small\hair 2pt
\pinlabel $i_\phi^U$ at 162 120
\pinlabel $G_\phi$ at 162 50
\pinlabel $B_\phi^U$ at 12 80 
\pinlabel $Z_\phi^U$ at 46 51
\pinlabel $V$ at 291 23 
%\pinlabel $i_\phi^U(W)$ at 195 14
\pinlabel $W$ at 89 80
%\pinlabel $i_\phi(W)$ at 261 92
\pinlabel $K$ at 222 60 
\pinlabel $K$ at 276 121 
\pinlabel $\Lambda$ at 244 100
\pinlabel $(p_1,x_2)$ at 320 82
\pinlabel $(x_1,p_2)$ at 233 130
\pinlabel $\graph \phi$ at 271 55
\endlabellist
\centering
\includegraphics[scale=1]{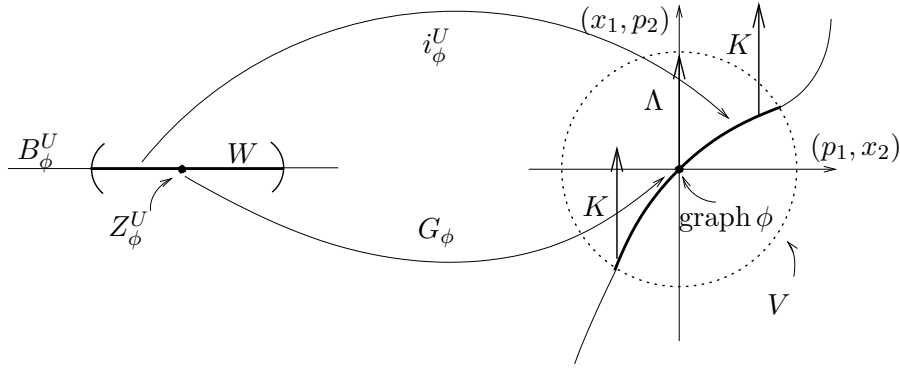}
\caption{Transversality condition and projectability on the
$(p_1,x_2)$--fibers.}
\label{fig:trans}
\end{figure}

\begin{proof}
In a local chart $U=\Cot U_1\times \Cot U_2$ of $\Cot M\times\Cot N$,
the bundle $G_\phi^*\Lambda$ is the restriction of
$$K:=\bigcup_{\big((p_1,x_1),(p_2,x_2)\big)\in\Cot U_1\times\Cot U_2}
\Bigg\{\big((0,v_1),(\mu_2,0)\big): v_1\in \R^k,\mu_2\in(\R^l)^*  \Bigg\}$$
to $\graph \phi = G_\phi(Z_\phi^U)$. 
The transversality condition tells us that 
the tangent space of $i_\phi^U(W)$ is transverse to K on $G_\phi(Z_\phi^U)$.
Now, by continuity, there exists a neighborhood $V$ of 
$G_\phi(Z_\phi^U)$ in $\Cot U_1\times\Cot U_2$ where
the tangent space $i_\phi^U(W)$ is transverse to $K$ on
$i_\phi^U(W)\cap V$. 
Observe that $K$ is the bundle transverse to the $p_1,x_2$ fibers.
This means that $i_\phi^U(W)\cap V$ is projectable on the
$p_1,x_2$ fibers and, thus, must be of the
form \eqref{belleforme}. 
This situation is illustrated in Figure \ref{fig:trans}.  
Considering that $i_\phi^U|_{Z_\phi^U}=G_\phi^U$, we get immediately that
$G(0,x_2) = \phi(x_2)$ and $H(0,x_2) = 0$.
\end{proof}

\begin{Proposition}\label{prop:cotbd1}
Let $\Psi:\Cot M\rightarrow \Cot N$ be
a symplectomorphism sending the zero section $Z_M$ to
the zero section $Z_N$. Then there exists a neighborhood
$V$ of $Z_M\times Z_N$ in $\Cot M\times\Cot N,$ 
and a transverse lagrangian germ
$i_\phi:B_\phi\hookrightarrow \overline{\Cot M}\times
\Cot N$ such that $$\graph \Psi\cap V = i_\phi(W),$$
where $W$ is a neighborhood of $Z_\phi$ in $B_\phi$.
\end{Proposition}

\begin{proof}
Let us prove the proposition in a local chart
$U= \Cot U_1\times\Cot U_2$ of $\Cot M\times\Cot N$.
Lemma \ref{lem:cotbd} tells us that $\Psi$ covers
a map $\phi:N\rightarrow M$. In the local chart $U$,
we have that:
$$B_\phi^U = \phi^*(\Cot U_1) = (\R^d)^*\times U_1.$$
Lemma \ref{lem:cotbd} gives a neighborhood $W$
of the zero section in $B_\phi^U$ and a neighborhood
$V$ of $U_1\times U_2$ in $\Cot U_1\times \Cot U_2$
where
$$
\graph\Psi\cap V = 
\Bigg\{
\bigg(\big(p_1,G(p_1,x_2)\big),\big(H(p_1,x_2),x_2\big)\bigg):
(p_1,x_2)\in W
\Bigg\}.
$$
such that $G(0,x_2) = \phi(x_2)$ and $H(0,x_2) = 0$.
Thus, there is a lagrangian germ $i_\phi^U:B_\phi^U\rightarrow
\overline{\Cot U_1}\times\Cot U_2$ around $G_\phi$ given by
$$i_\phi^U(p_1,x_2) = 
\Bigg(\big(p_1,G(p_1,x_2)\big),\big(H(p_1,x_2),x_2\big)\Bigg)
$$
such that $i_\phi^U(W) = \graph\Psi\cap V$.
Lemma \ref{lem:trans} tells us that $i_\phi^U$ is
transverse.
\end{proof}

\begin{Example}
Let $\phi$ be a map from $N$ to $M$ and
consider $d\phi^*:\Cot M\rightarrow\Cot N$ its 
cotangent lift. In a local chart $U=\Cot U_1\times\Cot U_2$,
the graph of $d\phi^*$ is:
$$\graph d\phi^* = 
\Bigg\{
\bigg(\big(p_1,\phi(x_2)\big),\big(d\phi^*(x_2)p_1,x_2\big)\bigg):
x_2\in U_2, \quad p_1\in(\R^d)^*
\Bigg\}.
$$
The induced transverse germ $i_\phi:B_\phi\hookrightarrow 
\overline{\Cot M}\times\Cot N$ is given by
$$i_\phi^U(p_1,x_2) = \big((p_1,\phi(x_2)),(d\phi^*(x_2)p_1,x_2)\big),$$
in the local chart $U$.
\end{Example}

The next proposition is a local converse of Proposition
\ref{prop:cotbd1}.

\begin{Proposition}\label{prop:germinv}
Let $\phi$ be a map from $N$ to $M$ and 
$i_\phi:B_\phi\hookrightarrow \overline{\Cot M}\times\Cot N$
be a transverse lagrangian germ around $G_\phi$.
If $\phi$ is invertible, then there exists  
a germ of symplectomorphism $\Psi$ which covers $\phi$ 
and such that:
$$\graph\Psi\cap V = i_\phi(W),$$
where $V$ is a neighborhood of $Z_M\times Z_N$ in $\Cot M\times
\Cot N$ and $W$ is a neighborhood of $Z_\phi$ in $B_\phi$.
\end{Proposition}

\begin{proof}
We prove the proposition in a local chart $U=\Cot U_1\times\Cot U_2$.
Lemma \ref{lem:trans} tells us that:
$$
i_\phi^U(W) =
\Bigg\{
\bigg(\big(p_1,G(p_1,x_2)\big),\big(H(p_1,x_2),x_2\big)\bigg)
:(p_1,x_2)\in W
\Bigg\}
$$
where $W$ is a neighborhood of $Z_\phi^U$ in $B_\phi^U$.
Now, consider the equation 
\begin{eqnarray}\label{eq:toinvert2}
G(p_1,x_2) & = & x_1.
\end{eqnarray}
Remark that $G(0,x_2) = \phi(x_2)$ and
$\nabla_{x_2}G(0,x_2) = \nabla \phi (x_2)$ which
is invertible as $\phi$ is a diffeomorphism. Then
the implicit function Theorem tells us that, for
$(p_1,x_2)$ in a neighborhood $W$ of
the zero section in $B_\phi^U$, 
we may invert equation \eqref{eq:toinvert2}, i.e.,
we may find a function $K$ such that
$$x_2 = K(p_1,x_1) \quad \textrm{s.t.}\quad
G(p_1,K(p_1,x_1)) = x_1.$$
Thus, we get that:
\begin{eqnarray*}
i_\phi^U(W) & = &
\Bigg\{
\bigg(\big(p_1,x_1\big),\big(H(p_1,K(p_1,x_1)),K(p_1,x_1):(p_1,x_1)\in T\big)
\bigg)
\Bigg\},
\end{eqnarray*}
where $T$ is a neighborhood of $U_1$ in $\Cot U_1$. Thus, setting
$$\Psi_U(p_1,x_1) = 
\Big(
H\big(p_1,K(p_1,x_1)\big),K(p_1,x_1)
\Big),
$$
and remarking that $\Psi_U(p_1,\phi(x)) = (0,x)$, one gets
a local description of a symplectomorphism germ $\Psi:\Cot M
\rightarrow \Cot N$ which covers $\phi$ and which sends a 
neighborhood of $M$ in $\Cot M$ to a neighborhood
of $N$ in $\Cot N$ preserving the bases.
\end{proof}

\section{Definition of the category}\label{sec:cat}

In this Section, we construct a new monoidal category, the 
cotangent microbundle category $\Mic$. 
Our goal is to extend (i.e. to replace maps by 
relations) the category of cotangent bundles so that 
the resulting ``category'' is a true category. The key
observation is the following. A morphism $\Psi:\Cot M\rightarrow \Cot N$
in the cotangent bundle category is a differentiable map
which satisfies the two following properties:
\begin{enumerate}
\item $\Psi$ is a symplectomorphism,
\item $\Psi$ preserves the zero sections (or cores).
\end{enumerate}
The idea is to reformulate these two properties in
terms of the graph of $\Psi$ so that they will still
make sense for general differentiable relations $L\subset
\overline{\Cot M}\times \Cot N$. It is well known that $\Psi$ is
a symplectomorphism if and only if its graph is a 
lagrangian submanifold of $\overline{\Cot M}\times\Cot N$. Now,
in the previous section, we have seen that asking that 
$\Psi$ preserves the cores is equivalent to ask
that its graph satisfies the transversality condition
of Definition \ref{def:trans}, which makes sense
for general lagrangian submanifolds of $\overline{\Cot M}
\times \Cot N$. However, this transversality condition
is a local condition. It concerns only the geometry
of the graph of $\Psi$ around a neighborhood of 
$$\graph \Psi\cap (Z_M\times Z_N) = \graph \phi,$$
in $\Cot M\times \Cot N$, where $\phi = \Psi_{|M}^{-1}$.
We are thus led to the following definition for the
morphism in $\Mic$.

\begin{comment}
This
category may be considered as a local version
of the extended symplectic category. 
However, in contrast to $\ExtSym$, $\Mic$ is
a true category.
The idea is to restrict ourselves, for the objects,
to  cotangent bundles of smooth
manifolds endowed with their canonical symplectic 
form. Instead of allowing all lagrangian submanifolds
of $\overline{\Cot M}\times\Cot N$ as morphisms between $\Cot M$
and $\Cot N$, we only take germs of lagrangian
submanifolds around graphs of maps from $N$ to
$M$ satisfying the transversality condition of Definition \ref{def:trans}.
Intuitively, these morphisms may be thought as deformations
of the conormal bundle of the graph of the underlying map $\phi$. 
\end{comment}
We keep the same notations as introduced in Section \ref{sec:trans}.

\begin{Definition} 
In $\Mic$, a morphism  from $\Cot M$ to $\Cot N$  is given
by a pair $(i_\phi, \phi)$ where $\phi$ is a map from $N$ to $M$
and $i_\phi$ is a transverse lagrangian germ as in Definition \ref{def:trans}.
%of lagrangian 
%embedding around $G_\phi$ such
%that $i_\phi$ is transverse to $G_\phi^*\Lambda$ on $Z_\phi$.
%We call such germs transverse germs for short.
\end{Definition}

\begin{Remark}
In the same way, we may define a ``micro'' version of
the extended symplectic category, the {\bf microsymplectic
category}, whose objects are pairs $(M,L)$ of a symplectic
manifold and a lagrangian submanifold $L\subset M$, its
core, and whose morphisms from $(M_1,L_1)$ to $(M_2,L_2)$
are pairs $(i_\phi,\phi)$ of
a smooth map $\phi:L_2\rightarrow L_1$ and a transverse
germ of lagrangian embeddings $$[i_\phi]:\phi^*(\Cot L_1)
\hookrightarrow \overline{M_1}\times M_2,$$
along $\phi$. In this context, we say that a germ
$[i_\phi]$ is transverse if for a representative $i_\phi$
(and hence any) of $[i_\phi]$ and for any identifications $\Psi_i$
of a neighborhood $U_i$ of $L_i$ in $M_i$ with a neighborhood
$V_i$ of $L_i$ in $\Cot L_i$, $i=1,2$, the induced germ
$$[i_\phi^\Psi]:\phi^*(\Cot L_1)\hookrightarrow \overline{\Cot L_1}\times\Cot L_2$$
is transverse in the sense of Definition \ref{def:trans}.
We will not pursue this point of view  this the article.
\end{Remark}

\begin{Example}\label{ex:first}
The category $\Mic$ has a distinguished object, 
the cotangent bundle of the one point
manifold $\neutral:= \Cot \{\star\}$. There is a unique morphism $\neutralmorph$
in $\morph(\neutral, \Cot M)$; it is given by  
$\neutralmorph = (i_M,pr)$ where $i_M$ is the inclusion of $M$ as
the zero section of $\Cot M$ and $pr$ is the projection of the whole
manifold $M$ onto $\star$.
\end{Example}

\begin{Example}
The base maps $\phi$ of morphisms $(i_\phi,\phi) 
\in \morph(\Cot M,\neutral)$ 
are indexed by points of $M$, namely $\phi:\{\star\}\rightarrow M$
sends the unique point $\star$ to a point $x$ of $M$. The transversality
condition in this context tells that images of
transverse germs of lagrangian embeddings
$i_\phi:B_{\phi}=\Cot{}_xM \hookrightarrow \Cot M\times \neutral$
are lagrangian submanifolds  
through $x\in M$ which are transverse to the zero section of $\Cot M$.
\end{Example}

\begin{Example}\label{ex:cotlift}
We may identify $\overline{\Cot M}\times\Cot N$ with
the cotangent bundle $\Cot(M\times N)$ via the Schwartz
transform: 
$$S
\big((p_1,x_1),(p_2,x_2)\big) =
\big((-p_1,p_2),(x_1,x_2)\big).$$
Let $\phi$ be a smooth map from the manifold $N$ to the manifold
$M$. The normal bundle $N^*(\graph \phi)$ of the graph of
$\phi$ in $\Cot(M\times N)$ induces, via the
Schwartz transform, a transverse germ of lagrangian embeddings:
$$d\phi^*:B_\phi\longrightarrow \overline{\Cot M}\times \Cot N.$$
We denote it by $d\phi^*$ as it comes from the cotangent lift
$d\phi^*:\Cot M\rightarrow\Cot N$ of $\phi$ if $\phi$ is invertible.
We call $(d\phi^*,\phi)$ the generalized cotangent lift
of $\phi$.
\end{Example}

\begin{Example}\label{ex:last}
For any cotangent bundle $\Cot M$, there is an identity morphism. It is given by 
$\id = (\Delta_{\Cot M},\id_M)\in\morph(\Cot M,\Cot M)$ where $\id_M$ is 
the identity map on $M$ and $\Delta_{\Cot M}$ the germ
induced by the diagonal in $\Cot M\times \Cot M$. 
\end{Example}
Consider a morphism
$(i_{\phi_1},\phi_1)$ from $\Cot M$ to $\Cot N$
and a morphism $(i_{\phi_2},\phi_2)$ from $\Cot N$ to $\Cot P$.
For two neighborhoods $N_1$ of $Z_{\phi_1}$ and $N_2$ of
$Z_{\phi_2}$, we may compose 
$i_{\phi_2}(N_2)\circ i_{\phi_1}(N_1)$ via composition of
canonical relations. 
The following proposition describes this composition.

\begin{Proposition}\label{prop:comp}
In the above notation, there exists 
a transverse germ of lagrangian embeddings
$i_{\phi_1\circ\phi_2}:B_{\phi_1\circ\phi_2}\hookrightarrow
\overline{\Cot M}\times \Cot P$ such that we may find
neighborhoods $N_k$ of $Z_{\phi_k}$, $k=1,2$,
and a neighborhood $N_3$ of $Z_{\phi_1\circ\phi_2}$  
for which:
$$i_{\phi_1\circ\phi_2}(N_3) = i_{\phi_2}(N_2)\circ i_{\phi_1}(N_1).$$
\end{Proposition}

\begin{proof}
We check the proposition in local coordinates. Let $U_1$, $U_2$
and $U_3$ be local charts of $M$, $N$ and $P$ respectively.
Denote by $V_1=\Cot U_1\times\Cot U_2$ and $V_2= \Cot U_2\times\Cot U_3$
the local charts of $\Cot M\times\Cot N$ and $\Cot N\times\Cot P$. 
In these charts, we have, thanks to Lemma \ref{lem:trans}, that:
\begin{eqnarray*}
i_{\phi_1}^{V_1}(N_1) & = & \Bigg\{ \bigg(\big(p_1,F(p_1,x_2)\big),\big(G(p_1,x_2),x_2\big)\bigg):(p_1,x_2)\in N_1 \Bigg\}\\
i_{\phi_2}^{V_2}(N_2) & = & \Bigg\{ \bigg(\big(p_2,H(p_2,x_3)\big),\big(L(p_2,x_3),x_3\big)\bigg):(p_2,x_3)\in N_2 \Bigg\}.
\end{eqnarray*}
The implicit function Theorem tells us that there exists a neighborhood $N_3$ of $Z_{\phi_1\circ\phi_2}^U$
in $B_{\phi_1\circ\phi_2}^U$ such that for $(p_1,x_3) \in N_3$, we can always find a
unique couple $(p_2,x_2)$ such that 
$$\big(G(p_1,x_2),x_2\big) = \big(p_2,H(p_2,x_3)\big).$$
Namely, consider the function:
\[
I(p_1,x_3, p_2, x_2) = \left(\begin{array}c
 p_2- G(p_1, x_2)\\
 x_2- H(p_2,x_3)
\end{array}\right).
\]
Thanks to the fact that $G(0,x_1) =  0$ and
$H(0,x_3) = \phi_2(x_3)$, we get 
$$I(0,x_3,0,\phi_2(x_3)) = 0.$$
Moreover, the Jacobi matrix of $I$ at this point 
\[
\frac{\partial I}{\partial(p_2, x_2)}\big((0,x_3,0,\phi(x_3)\big) = \left(\begin{array}{cc}
\id & 0\\
-\nabla_p H(0,x_3) & \id\end{array}\right)
\]
is invertible. This shows that there exists a neighborhood 
$N_3$ of the zero section in $B_{\phi_1\circ\phi_2}^U$ 
and a unique solution $p_2 = U(p_1,x_3)$ and $x_2 = V(p_1,x_3)$
such that:
\begin{gather}\label{eq:implfct}U(p_1,x_3) = H(V(p_1,x_3),x_3),
        \quad V(p_1,x_3) = G(p_1,U(p_1,x_3)),
\end{gather}
for $(p_1,x_3) \in N_3$.
Then we have that composition of 
canonical relations yields:
\begin{eqnarray*}
i_{\phi_2}^{V_2}(N_2) \circ i_{\phi_1}^{V_1}(N_1) & = & 
\Bigg\{ \bigg(\big(p_1,R(p_1,x_3))\big),\big(T(p_1,x_3),x_3\big)\bigg):(p_1,x_3)\in N_3 \Bigg\},\\
\end{eqnarray*}
where $R(p_1,x_3) = F(p_1,U(p_1,x_3))$ and $T(p_1,x_3) =  L(V(p_1,x_3),x_3)$.
Setting $p_1 = 0$ in \eqref{eq:implfct},
we get that $U(0,x_3) = \phi_2(x_3)$ and $V(0,x_3) = 0$ and then
$R(0,x_3) = \phi_1\circ\phi_2(x_3)$ and $T(0,x_3) = 0$. 
In conclusion, $i_{\phi_2}^{V_2}(N_2) \circ i_{\phi_1}^{V_1}(N_1)$
is a true lagrangian submanifold and defines a germ of lagrangian embeddings
$$i_{\phi_1\circ\phi_2}:B_{\phi_1\circ\phi2}\hookrightarrow \overline{\Cot M}\times\Cot N.$$
around $G_{\phi_1\circ\phi_2}$. By Lemma \ref{lem:trans}, this germ is transverse
to $G^*\Lambda$.
\end{proof}

\begin{Definition}
Let $(i_{\phi_1},\phi_1)\in\morph(\Cot M,\Cot N)$ and
$(i_{\phi_2},\phi_2)\in\morph(\Cot N,\Cot P)$ be two morphisms
in $\Mic$.  We define the composition between them 
by $$ (i_{\phi_2},\phi_2)\circ (i_{\phi_1},\phi_1) 
:= (i_{\phi_1\circ\phi_2},\phi_1\circ\phi_2),$$
where $i_{\phi_1\circ\phi_2}$ is the germ
obtained  in Proposition \ref{prop:comp} by the usual composition
of canonical relations and $\phi_1\circ\phi_2$ is the usual 
composition of maps. 
\end{Definition}

\begin{Example}
Suppose we have a map $\phi_1$ from $N$ to $M$ and a map
 $\phi_2$ from $M$ to $Q$. Consider the generalized cotangent lifts
$(d\phi_1^*,\phi_1)$ and $(d\phi_2^*,\phi_2)$ as 
defined in Example \ref{ex:cotlift}. Then we have that
$(d\phi_2^*,\phi_2)\circ(d\phi_1^*,\phi_1) = 
\big(d(\phi_1\circ \phi_2)^*,\phi_1\circ \phi_2\big).$
\end{Example}

We may also define a bifunctor:
$$\otimes:\Mic\times\Mic\longrightarrow\Mic$$
in the following way. Take two cotangent bundles $\Cot M$,$\Cot N$.
We define the product between objects as $\Cot M\otimes \Cot N = \Cot(M\times N).$
Take two morphisms $(i_{\phi_1},\phi_1)\in\morph(\Cot M,\Cot N)$
and $(i_{\phi_2},\phi_2)\in\morph(\Cot P,\Cot Q)$. The product between
morphisms is given by
$(i_{\phi_1},\phi_1)\otimes (i_{\phi_2},\phi_2) = (i_{\phi_1}\times i_{\phi_2},\phi_1\times\phi_2).$
The bifunctoriality of $\Mic$ follows trivially from the bifunctoriality of
the Cartesian product on sets. 
Let us summarize the results obtained so far in the following theorem.

\begin{Theorem}
$\Mic$ is a monoidal category.
\end{Theorem}

%\begin{proof}
%Proposition \ref{prop:trans}, that the composition between morphisms
%is alway well defined. The associativity of morphism composition follows trivially from
%the associativity of the composition of maps and canonical relations.
%\end{proof}

\section{Generating functions}\label{sec:gen}
In this Section, we describe morphisms $(i_\phi,\phi)$ from
$\Cot M$ to $\Cot N$ in local charts in terms of a single
function: the generating function $S$ of the lagrangian
embedding $i_\phi$. We derive then a composition formula
for these generating functions which represents the composition
of morphisms. At last, we see how these generating functions
behave under change of coordinates.

Observe first that, for any manifold $M$, we may always find a system 
of star-shaped charts $\{U_\alpha\}_{\alpha\in A}$ which
covers $M$. In the sequel, we always assume that the charts
of the base manifolds are of this sort.
In particular, we consider the induced charts of $\Cot M\times\Cot N$ of
the type $U=\Cot U_1\times\Cot U_2$ where
$U_1$ and $U_2$ are star-shaped charts of 
$M$ and $N$ respectively.
Now, if $i_\phi:B_\phi \hookrightarrow \overline{\Cot M}\times
\Cot N$ is a transverse germ of lagrangian embeddings around
$G_\phi$ , Lemma \ref{lem:trans} tells us that 
there exists a neighborhood $W$ of $Z_\phi^U$ such
that:
$$
i_\phi^U(W) = \Bigg\{
\bigg(\big(p_1,G(p_1,x_2)\big),
\big(H(p_1,x_2),x_2\big)\bigg): (p_1,x_2)\in W
\Bigg\}.$$
The fact that $i_{\phi}^U(W)$ is a lagrangian 
submanifold and that $U$ is topologically trivial
implies that there exists a function
$S_U:W\rightarrow\R$ such that
$$G(p_1,x_2) = \nabla_{p_1}S_U(p_1,x_2) \quad\textrm{ and }\quad
H(p_1,x_2) = \nabla_{x_2}S_U(p_1,x_2).$$
The fact that $i_\phi^U|_{Z_\phi^U}= G_\phi^U$ imposes that:
\begin{eqnarray}
\label{cond}
\nabla_{x_2}S_U(0,x_2) = 0
\quad\textrm{ and }\quad
\nabla_{p_1}S_U(0,x_2) = \phi(x_2).
\end{eqnarray} 
This implies that $S_U(0,x_2)$ is equal to a constant. We 
may normalize $S_U$ by choosing this constant
to be zero. From now on, we consider only normalized
generating functions.

\begin{Definition}
We call $S_U$ as above the generating
function of the morphism $(i_\phi,\phi)\in \morph(\Cot M,\Cot N)$
in the local chart $U=\Cot U_1\times\Cot U_2$.
Notice that $S_U$ may be considered as a germ of
 smooth functions on $B_\phi^U$ around $Z_\phi^U$.
\end{Definition}
We provide now the generating functions of the morphisms
given in Examples \ref{ex:first} to \ref{ex:last}.

\begin{Example}\label{ex:genneutral}
Consider the unique morphism $(i_M,pr)$ of $\morph(\neutral,\Cot M)$.
Then, in a local chart, $B_{pr}= M$. The unique normalized
function $\neutralmorph:M\rightarrow \R$  is
the zero function, $\neutralmorph(x) = 0$.
\end{Example}

\begin{Example}
Take a morphism $(i_\phi,\phi)\in\morph(\Cot M,\neutral)$.
In this case, $B_{\phi} = \Cot{}_x M$ where $x = \phi(\star)$.
Then the generating function is a germ of
functions $F:\Cot{}_xM\rightarrow\R$ around $0$ such that
$F(0) = 0$ and $\nabla F(0) = x$. Let $V=\Cot U$
be local chart of $\Cot M$. Then
$$i_{\phi}^V(N) = \Big\{\big(p,\nabla F(p)\big)\times \{\star\}:p\in \Cot{}_xM\Big\}.$$ 
\end{Example}

\begin{Example}\label{ex:genlift}
Consider $(d\phi^*,\phi)\in\morph(\Cot M,\Cot N)$. 
In a local chart $U:=\Cot U_1\times\Cot U_2$, 
we have that $B_\phi^U=(\R^m)^*\times U_2$ and the generating
function of $d\phi^*$ is the function $G_\phi:B_\phi^U\rightarrow\R$
given by $G_\phi(p_1,x_2) = \langle\phi(x_2),p_1\rangle$.
\end{Example}

\begin{Example}\label{ex:genid} As a special instance of
Example \ref{ex:genlift}, the generating
function of the identity morphism $\id=(\Delta_{\Cot M},\id_M)
\in\morph(\Cot M,\Cot M)$ is $G_\Delta(p_1,x_2) = p_1x_2$.

\end{Example}
Let us see how composition of morphisms reflects locally on
their generating functions. For that consider
some local charts $U_1$, $U_2$ and $U_3$  
of respectively $M$, $N$ and $P$. Let $(i_{\phi_1},\phi_1)
\in\morph(\Cot M,\Cot N)$ and $(i_{\phi_2},\phi_2)
\in\morph(\Cot N,\Cot P)$. We denote by $G$ and $F$
the generating functions of $(i_{\phi_1},\phi_1)$ and
$(i_{\phi_2},\phi_2)$ in the local charts $V_1 = \Cot U_1\times\Cot U_2$
and $V_2 = \Cot U_2\times\Cot U_3$. Let also be
$V_3 = \Cot U_1\times\Cot U_3$ the local chart
of $\Cot M\times\Cot P$.

\begin{Definition}
Let $f\in C^\infty(\R^k)$ be a function which has only one
critical point on $\R^k$. We denote by
$\stat{x}\{f\}$ the value of $f$ at its critical point $x_0$, i.e.,
at the point $x_0$ such that $\nabla_xf(x_0) = 0.$ If $f$ depends
on the variables $x,y$, we denote by $\stat{x}\{f\} (y)$ the
function depending on $y$ defined by $f(x_0(y),y)$ where
$x_0(y)$ is the implicit function solution of the equation
$\nabla_x f(x_0(y),y) = 0$. 
\end{Definition}

\begin{Lemma}\label{lem:critpt} There is a neighborhood $N$ of the zero
section of $B_{\phi_1\circ\phi_2}^{V_3}$ such that,
for all $(p_1,x_3)\in N$, the function
$$H(\bar p,\bar x) := F(\bar p, x_3)+G(p_1,\bar x)-\bar p\bar x$$
has only one critical point with respect to the $\bar p,\bar x$ 
variables.
\end{Lemma}

\begin{proof}
The critical points $\bar p$ and $\bar x$ are the solution of the following system
of implicit equations:
$\bar p = \nabla_{x_2} G(p_1,\bar x)$ and $\bar x = \nabla_{p_2}F(\bar p,x_3).$
The implicit function theorem tells us that this system
has always a unique solution for small enough $p_1$. Namely, set
\[
H(p_1,x_3,\bar p,\bar x) = \left(\begin{array}c
\bar p-\nabla_{x_2} G(p_1,\bar x)\\
\bar x-\nabla_{p_2} F(\bar p,x_3)
\end{array}\right).
\]
Thanks to the fact that $G(0,x_1) =  0$ and
$\nabla_{p_2}F(0,x_3) = \phi_2(x_3)$, we get 
$$H(0,x_3,0,\phi_2(x_3)) = 0$$ which means
that for $p_1=0$, the critical points are $\bar p = 0$ and 
$\bar x = \phi_2(x_3)$. Moreover, the Jacobi matrix of $H$ at this point
with respect to the $\bar p,\bar x$ variables
\[
\frac{\partial H}{\partial(\bar p,\bar x)}\big((0,x_3,0,\phi(x_3)\big) = \left(\begin{array}{cc}
\id & 0\\
-\nabla_p\nabla_pF(0,x_3) & \id\end{array}\right)
\]
is invertible. This shows that, for  $(p_1,x_3)$ in
a neighborhood $N$ of the zero section in 
$B_{\phi_1\circ\phi_3}^{V_3}$, $H$ always possesses 
unique critical points $\bar p$ and $\bar x$.  
\end{proof}

\begin{Definition}\label{def:gencomp} Let $F$ and $G$ be as above. We define
the composition of generating function as:
\begin{eqnarray}\label{CompositionFormula}
F\circ G (p_1,x_3) & := &
\stat{\bar p,\bar x}
\Big\{F(\bar p, x_3)+G(p_1,\bar x)-\bar p\bar x \Big\}.
\end{eqnarray}
Note that Lemma \ref{lem:critpt} guarantees that
the composition is well-defined.
\end{Definition}

\begin{Lemma}\label{lem:der}
In the above notation, we have that
$$\nabla_p (F\circ G)(p_1,x_3)  = \nabla_pG(p_1, \bar x)$$
$$\nabla_x (F\circ G)(p_1,x_3)  = \nabla_pF(\bar p, x_3)$$
where $\bar p$ and $\bar x$ are solutions of the implicit system:
\begin{eqnarray}
\label{critpts1} \bar p & = & \nabla_x G(p_1,\bar x),\\
\label{critpts2} \bar x & = & \nabla_p F(\bar p,x_3).
\end{eqnarray}
\end{Lemma}

\begin{proof}
From Definition \ref{def:gencomp}, we have that
$$F\circ G (p_1,x_3)  := F(\bar p, x_3)+G(p_1,\bar x)-\bar p\bar x,$$
where $\bar p$ and $\bar x$ is the unique solution of
the system \eqref{critpts1}--\eqref{critpts2}.
Deriving $F\circ G$ with respect to $p_1$, we get
that:
$$\nabla_p (F\circ G)(p_1,x_3) =
\nabla_p G(p_1,\bar x) + 
\nabla_p F(\bar p,x_3)\frac{d\bar p}{dp_1}
+ \nabla_x G(p_1,\bar x)\frac{d\bar x}{dp_1}  
- \frac{d(\bar p\bar x)}{dp_1}.
$$
Noticing that,
$$\frac{d(\bar p\bar x)}{p_1} = 
\nabla_p F(\bar p,x_3)\frac{d\bar p}{dp_1}
+ \nabla_x G(p_1,\bar x)\frac{d\bar x}{dp_1},
$$
we get that:
$$\nabla_p (F\circ G)(p_1,x_3)  = \nabla_pG(p_1, \bar x).$$
Similarly, we get that:
$$\nabla_x (F\circ G)(p_1,x_3)  = \nabla_xF(\bar p, x_3).$$
\end{proof}

\begin{Lemma} Let $F$ and $G$ be as above, then we have that:
$$F\circ G(0,x) = 0\quad\textrm{ and }\quad \nabla_{p_1}F\circ G(0,x) = \phi_1\circ\phi_2(x).$$
\end{Lemma}

\begin{proof}
The critical points are given by the equations
$\bar p = \nabla_x G(p_1,\bar x)$ and 
$\bar x = \nabla_p F(\bar p, x_3)$. If $p_1=0$, we
get that $\bar p= 0$ and $\bar x = \phi_2(x_3)$.
Thus, we have immediately that  $F\circ G(0,x_3) = 0.$
Lemma \ref{lem:der} tells us that:
$\nabla_{p_1} (F\circ G(p_1,x_3))  = \nabla_{p_1}G(p_1,\bar x)$
and thus, we have that:
$$\nabla_{p_1} (F\circ G(0,x_3))  
= \nabla_{p_1}G(0,\phi_2(x_3)) = \phi_1\circ\phi_2(x_3).$$
\end{proof}

\begin{Proposition} Let $F$ and $G$ be as above, then
$F\circ G$ is the generating function of 
$(i_{\phi_2},\phi_2)\circ (i_{\phi_1},\phi_1)$
in the local chart $\Cot U_1\times\Cot U_3$.
\end{Proposition}

\begin{proof}
In the local charts $V_1$ and $V_2$, we have that:
\begin{eqnarray*}
i_{\phi_1}(N_G) & = & \Big \{ \Big(\big(p_1, \nabla_p {G}(p_1,x_2)\big), \big(\nabla_x{G}(p_1,x_2), x_2\big) \Big): (p_1,x_2)\in N_{G}\Big\}\\
i_{\phi_2}(N_F) & = & \Big \{ \Big(\big(p_2, \nabla_p F(p_2,x_3)\big), \big(\nabla_xF(p_2,x_3), x_3 \big) \Big): (p_2,x_3)\in N_F\Big\}
\end{eqnarray*}
where $N_F$ and $N_G$  are the neighborhood of the zero section
in respectively $B_{\phi_1}^{V_1}$ and $B_{\phi_2}^{V_2}$.
The composition of canonical relations yields:
\begin{eqnarray*}
i_{\phi_2}(N_F)\circ i_{\phi_1}(N_G) 
                & = &  \Big \{ \Big (\big( p_1, \nabla_p G(p_1,x_2)\big) ,\big( \nabla_x F(p_2,x_3),x_3\big)\Big):\\
                &   & :\quad x_2 = \nabla_p F(p_2,x_3),\quad p_2 = \nabla_x G(p_1,x_2), \quad(p_1,x_3)\in N \Big \}\\
\end{eqnarray*}
where $N$ is a neighborhood of the zero section
in $B_{\phi_1\circ\phi_2}^{V_3}$ where the system,
\begin{eqnarray}
 p_2 & = & \nabla_x G(p_1,x_2),\\
 x_2 & = & \nabla_p F(p_2,x_3),
\end{eqnarray}
has a unique solution $(p_2,x_2)$ for $(p_1,x_3)\in N$.
Lemma \ref{lem:critpt} tells us that $F\circ G$ is exactly defined on $N$
and induces a lagrangian germ described by
\begin{eqnarray*}
i_{F\circ G}(N) & = & \Big \{ \Big(\big(p_1, \nabla_p (F\circ G)(p_1,x_3)\big), \big(\nabla_x (F\circ G)(p_1,x_3),x_3 \big) :(p_1,x_3)\in N\Big\}.
\end{eqnarray*}
An inspection of Lemma \ref{lem:der} shows that $F\circ G$ is the generating function 
of $(i_{\phi_1},\phi_1)\circ(i_{\phi_2},\phi_2)$ in
the local chart $\Cot U_1\times\Cot U_3$.
\end{proof}

Suppose we have a morphism $T=(i_\phi,\phi)$ from $\Cot M$
to $\Cot N$ and a morphism $L=(i_\psi,\psi)$ from $\Cot P$ 
to $\Cot Q$. The tensor product $T\otimes L$ is then
a morphism from $\Cot(M\times P)$ to
$\Cot(N\times Q)$. Let $p_1,x_1,p_2,x_2$ be local coordinates
on $\Cot M\times \Cot N$ and $\bar p_1,\bar x_1,\bar p_2,\bar x_2$
local coordinates on $\Cot P\times \Cot Q$ and let
$F$ and $G$ be the generating functions of $T$ and
$L$.
The generating function of $T\otimes L$ in these charts
is a germ of a smooth function 
$F\otimes G$ on $B_{\phi\times\psi} = B_\phi\times B_\psi$
around the zero section.
Note that the induced local coordinates on $B_{\phi\times\psi}$
are $p_1,\bar p_1,x_2,\bar x_2$.
The following lemma gives us the form of $F\otimes G$.

\begin{Lemma}\label{lem:oplus}In the above notation,
the generating function $F\otimes G$ of $T\otimes L$ 
 is a germ of smooth functions
$F\otimes G:B_{\phi\times\psi}\rightarrow \R$
around the zero section given
by:
$$F\otimes G(p_1,\bar p_1,x_2,\bar x_2) := F(p_1,x_2)+G(\bar p_1,\bar x_2).$$
\end{Lemma}
\begin{proof}
One sees that directly on the graph of $T\otimes L$ written
in the local coordinates as above:
$$
\Bigg\{
\bigg(p_1,\bar p_1,\nabla_{p_1}F(p_1,x_2),
\nabla_{\bar p_1}G(\bar p_1,\bar x_2)\bigg),
\bigg(\nabla_{x_2}F(p_1,x_2),\nabla_{\bar x_2}G(\bar p_1,\bar x_2),x_2,\bar x_2\bigg).
\Bigg\}
$$
 
\end{proof}

Before ending this Section, we describe how the generating functions 
behave locally when changing coordinates.
Suppose we have two local charts $W_\alpha = \Cot U_\alpha\times\Cot V_\alpha$
and $W_\beta= \Cot U_\beta\times\Cot V_\beta$ of 
$\Cot M\times\Cot N$. Let us denotes by respectively
$S_\alpha$ and $S_\beta$ the generating functions of
the local restriction $(i_{\phi_\alpha},\phi_\alpha)$ and
$(i_{\phi_\beta},\phi_\beta)$ of 
a morphisms $(i_\phi,\phi)\in\morph(\Cot M,\Cot N)$
in these local charts.
If $g:U_\alpha\rightarrow U_\beta$ 
and $h:V_\beta\rightarrow V_\alpha$ 
are the changes of coordinates on the base manifolds,
then $(dg^*,g)\in\morph(\Cot U_\beta,\Cot U_\alpha)$
and $(dh^*,h)\in\morph(\Cot V_\alpha,\Cot V_\beta)$ (see
Example \ref{ex:cotlift}). Let us denote by
$p_1,x_1,p_2,x_2$ the local coordinates on 
$W_\alpha$ and by $\bar p_1,\bar x_1,\bar p_2,\bar x_2$
the local coordinates on $W_\beta$. In these coordinates,
the generating function of $(dg^*,g)$ is 
$G(\bar p_1,x_1) = g(x_1)\bar p_1$ and the generating
function of $(dh^*,h)$ is $H(p_2,\bar x_2) = h(\bar x_2) p_2$.

\begin{Lemma}\label{lem:compform} In the above notation, we have that:
$$H\circ S_\alpha \circ G(\bar p_1,\bar x_2) 
= S_\alpha\circ G(\bar p_1, h(\bar x_2)).$$
\end{Lemma}

\begin{proof}
By definition, we have that:
$$ H\circ (S_\alpha\circ G)(\bar p_1,\bar x_2) =
H(\tilde p,\bar x_2) + (S_\alpha\circ G)(\bar p_1,\tilde x) - \tilde p\tilde x,$$
where the critical point computation yields: 
$$\tilde p = \nabla_x (S_\alpha\circ G)(\bar p_1,\tilde x)\quad \textrm{ and }\quad 
  \tilde x = \nabla_p H(\tilde p, \bar x_2).$$
Computing $S_\alpha\circ G( \bar p_1,\tilde x)$, we get:
$$ S_\alpha\circ G( \bar p_1,\tilde x) =
 S_\alpha(\hat p,\tilde x) + G(\bar p_1, \hat x) - \hat p\hat x,$$
where the critical points are given by:
$$ \hat p = \nabla_x G(\bar p_1, \hat x)
\quad \textrm{and} \quad
\hat x = \nabla_p S_\alpha(\hat p,\tilde x).
$$
Remarking that $\tilde x = h(\bar x_2)$ and that
$\tilde  p\tilde  x = H(\tilde p,\bar x_2)$, we get that:
$$ H\circ (S_\alpha\circ G)(\bar p_1,\bar x_2) =
S_\alpha(\hat p,h(\bar x_2)) + G(\bar p_1, \hat x) - \hat p\hat x,$$
where 
$$ \hat p = \nabla_x G(\bar p_1, \hat x)\quad \textrm{ and }\quad
\hat x = \nabla_p S_\alpha(\hat p,h(\bar x_2)).$$
Thus, we get that:
$$H\circ S_\alpha \circ G(\bar p_1,\bar x_2) =
\stat{\hat p,\hat x}
\Big\{
S_\alpha(\hat p,h(\bar x_2)) + G(\bar p_1, \hat x) - \hat p\hat x,
\Big\},
$$
which ends the proof.
\end{proof}

Suppose that we are given
a collection of morphisms $(i_{\phi_\gamma},\phi_\gamma)
\in\morph(\Cot U_\gamma,\Cot V_\gamma)$ on local 
charts $\{\Cot U_\gamma\times \Cot V_\gamma\}_{\gamma\in A}$
of $\Cot M\times \Cot N$ whose generating functions are
denoted by $S_\gamma$. Suppose further that the
$\phi_\gamma:V_\gamma\rightarrow U_\gamma$ are the
restrictions of a global morphism $\phi:N\rightarrow M$.
The following proposition tells us when this collection 
$C:=\{(i_{\phi_\gamma},\phi_\gamma)\}_{\gamma\in A}$ of local
morphisms comes from a global morphism $(i_\phi,\phi)\in
\morph(\Cot M,\Cot N)$.

\begin{Proposition}\label{prop:coordchange}
Let $C:=\{(i_{\phi_\gamma},\phi_\gamma)\}_{\gamma\in A}$ 
be a collection of local morphisms corresponding to local
charts $\Cot U_\gamma\times\Cot V_\gamma$
of $\Cot M\times\Cot N$ as above.
The following statements are equivalents:
\begin{itemize}
\item[(1)] The collection $C$ comes from the restrictions of
a global morphism $(i_\phi,\phi)\in \morph(\Cot M,\Cot N)$ to
the local charts.

\item[(2)] For any two morphisms 
 $(i_{\phi_\alpha},\phi_\alpha), (i_{\phi_\beta},\phi_\beta)\in C$
we have, on overlapping domains, that:
$$(i_{\phi_\beta},\phi_\beta)
= (dh^*,h)\circ (i_{\phi_\alpha},\phi_\alpha)\circ (dg^*,g),$$
where $g:U_\alpha\rightarrow U_\beta$ and 
$h:V_\beta\rightarrow V_\alpha$ are the change of
coordinates.
\item[(3)] 
For any two morphisms 
 $(i_{\phi_\alpha},\phi_\alpha), (i_{\phi_\beta},\phi_\beta)\in C$
we have, on overlapping domains, that:
$$S_\beta = H\circ S_\alpha\circ G,$$
where $S_\alpha$ and $S_\beta$ are the generating
function of the local morphisms and $H$ and $G$ are
the generating functions of respectively
$(dh^*,h)$ and $(dg^*,g)$.
\end{itemize}
\end{Proposition}

\begin{proof}
By definition, (2) and (3) are equivalent. We show here
that (3) and (1) are also equivalent.
To simplify the notation, we suppose that 
$\Cot U_\alpha\times \Cot V_\alpha$ and $\Cot U_\beta
\times \Cot V_\beta$ describe the same open
subset of $\Cot M\times\Cot N$.
The graph of $(i_{\phi_\alpha},\phi_\alpha)$ in 
$\Cot U_\alpha\times \Cot V_\alpha$ is given by:
$$
L_\alpha = \Big \{ \Big(\big(p_1, \nabla_p S_\alpha(p_1,x_2)\big),
\big(\nabla_x S_\alpha(p_1,x_2), x_2\big) \Big): (p_1,x_2)\in N_\alpha\Big\}\\
$$
where $N_\alpha$ is a neighborhood of the zero section in $B_{\phi}^\alpha$.
Similarly, the graph of $(i_{\phi_\beta},\phi_\beta)$ in $\Cot U_\beta
\times \Cot V_\beta$ is given by:
$$
L_\beta = \Big \{ \Big(\big(\bar p_1, \nabla_{\bar p} S_\beta(\bar p_1,\bar x_2)\big),
\big(\nabla_{\bar x}S_\beta(\bar p_1,\bar x_2), \bar x_2 \big) \Big):
 (\bar p_1,\bar x_2)\in N_\beta\Big\}
$$
where $N_\beta$ is a neighborhood of the zero section in
$B_\phi^\beta$.
Now, $L_\alpha$ and $L_\beta$ describe the same
submanifold of $\Cot M\times \Cot N$ iff 
\begin{gather*}
p_1 = dg_{\beta\alpha}^*\big(\nabla_p S_\alpha(p_1,x_2)\big)\bar p_1,
\quad \nabla_p S_\alpha(p_1,x_2) = g_{\alpha\beta}
\big( \nabla_{\bar p} S_\beta(\bar p_1,\bar x_2) \big)\\
\nabla_x S_\alpha(p_1,x_2) = dh_{\beta\alpha}^*(x_2) 
\big(\nabla_{\bar x} S_\beta(\bar p_1,\bar x_2)\big),
\quad x_2 = h_{\alpha\beta}(\bar x_2).
\end{gather*}
This is equivalent to have that:
\begin{eqnarray}
\label{eq:coord1}\nabla_{\bar x}S_\beta(\bar p_1,\bar x_2)
& = & dh_{\alpha\beta}^*(\bar x_2)
\big(\nabla_x S_\alpha(\hat p, h_{\alpha\beta}(\bar x_2))\big)\\
\label{eq:coord2}\nabla_{\bar p}S_\beta (\bar p_1,\bar x_2) & = & g_{\beta\alpha}(\hat x)
\end{eqnarray}
where
$ \hat x = \nabla_p S_\alpha(\hat p,h_{\alpha\beta} (\bar x_2))$ and
$\hat p = dg_{\beta\alpha}^*(\hat x)\bar p_1.  $
Now, thanks to Lemma \ref{lem:compform}, we have that:
\begin{eqnarray}\label{uuuu}
H\circ S_\alpha \circ G(\bar p_1,\bar x_2)
& = & S_\alpha(\hat p, h_{\alpha\beta}(\bar x_2)) + 
G(\bar p_1,\hat x) -\hat p\hat x,
\end{eqnarray}
where we also have that
$ \hat x = \nabla_p S_\alpha(\hat p,h_{\alpha\beta} (\bar x_2))$ and
$\hat p = dg_{\beta\alpha}^*(\hat x)\bar p_1.$
Applying Lemma \ref{lem:der} to $H\circ S_\alpha \circ G$
we get equations \eqref{eq:coord1}--\eqref{eq:coord2}.
Thus, this shows that (3) implies (1). On the other hand,
(1) implies that the derivative of the generating function $S_\beta$
have the form given by equations \eqref{eq:coord1}--\eqref{eq:coord2}.
The only normalized generating function which has these
derivatives is $H\circ S_\alpha\circ G$.
\end{proof}

\section{The Poisson functor}\label{sec:gpd}
This Section is devoted to showing that a monoid structure
on an object $\Cot M$ of the cotangent microbundle category induces
a Poisson structure on the base manifold $M$ together with
a local symplectic groupoid $(s,t):\Cot M\rightrightarrows M$
integrating it. The description of both the Poisson structure
and the local symplectic groupoid are given explicitly in
terms of the generating function of transverse lagrangian
germs. We also prove that morphisms of monoid structures
produce Poisson morphims on the base.
This yields, in particular, a contravariant functor
$$\Deq:\Mon\Mic\longrightarrow\Pois,$$
form the category $\Mon\Mic$ of monoid objects and
monoid maps in $\Mic$ to the category $\Pois$ of Poisson
manifolds and Poisson maps.

\begin{Definition}
In a monoidal category $C$ with neutral object $\neutral$,
a monoid is a triple $(M,\mu,\neutralmorph)$
made of an object $M$, a morphism $\mu\in\morph(M^{\otimes2},M)$
called the product and a morphism $\neutralmorph\in\morph(\neutral,M)$
called the unit. These morphisms should satisfy the two
following relations:
\begin{gather}
\label{ass:ass}\mu\circ(\mu\otimes\id) = \mu\circ(\id\otimes\mu)\\
\label{ass:neut}\mu\circ(\neutralmorph\otimes\id) 
= \mu\circ(\id\otimes\neutralmorph) = \id.
\end{gather}
We call the couple $(\mu,\neutralmorph)$ a monoid structure on $M$.
\end{Definition}

\begin{Definition}
Let $\mathcal C$ be a monoidal category and let 
$(M,\mu_M,\neutralmorph_M)$ and $(N,\mu_N,\neutralmorph_N)$
be two monoids in $C$. We say that a morphism
$T:M\rightarrow N$ is a monoid morphism
if $T\circ \mu_M = \mu_N\circ (T\otimes T)$
and $T(e_M) = e_N$.
\end{Definition}

It is easy to see that the monoid object in a monoidal
category $\mathcal C$ together with their monoid
morphisms form a category, which we denote
by $\Mon{\mathcal C}$.

\begin{Example}
A monoid $(V,\mu,\neutralmorph)$ in
the category of complex vector spaces is a usual unital
algebra. The morphism $\mu:V\otimes V\rightarrow V$ is
the associative product and the unit morphism 
$\neutralmorph:\C\rightarrow V$ is given by $\neutralmorph(\lambda)
= \lambda\cdot 1$ where $1$ is the unit of the algebra. 
\end{Example}

The following Proposition tells us that a monoid 
 $(\Cot M, \mu, e)$ in $\Mic$ is completely
determined by its product $\mu\in\morph(\Cot M^{\otimes 2},\Cot M)$
whose base map must be the diagonal map
$\Delta:M\rightarrow M\times M$.

\begin{Proposition}\label{lem:rigidity} Let $(\Cot M,\mu,\neutralmorph)$ 
be a monoid in $\Mic$. Then the unit morphism is
the unique morphism of $\morph(\neutral, \Cot M)$, i.e.,
$\neutralmorph  =  (i_M,pr)$, and the product 
$\mu\in\morph(\Cot M^{\otimes2},\Cot M)$ is of the form
$\mu  =  (i_\Delta,\Delta)$
where $\Delta:M\rightarrow M\times M$ is the diagonal map
$\Delta(x) = (x,x).$
\end{Proposition}

\begin{proof}
As $\morph(\neutral,\Cot M)$ possesses only one element given
by $(i_M,pr)$, this imposes that $\neutralmorph = (i_M,pr)$.
Suppose now that $\mu  =  (i_\phi,\phi)$
satisfies \eqref{ass:neut}, i.e.,
\begin{eqnarray*}
 (\Delta_{\Cot M},\id_M) & = &  
\big(i_\phi\circ(\Delta_{\Cot M}\times i_M),(\id_M\times pr)\circ\phi\big)\\
& = & \big(i_\phi\circ(i_M\times\Delta_{\Cot M}),(pr\times\id_M)\circ\phi\big).
\end{eqnarray*}
If we set $\phi(x) = (\phi_1(x),\phi_2(x))$, then
$(\id_M\times pr)\circ\phi = \id_M$
translates into $\phi_1(x) = x$ and $(pr\times\id_M)\circ\phi=\id$ 
into $\phi_2(x) = x$. Thus, $\phi(x) = (x,x)$.
\end{proof}

Proposition \ref{lem:rigidity} tells us that monoid structures on an object
$\Cot M$ in $\Mic$ are entirely determined by germs
of lagrangian embedding, $$i_\Delta:B_\Delta\hookrightarrow 
\overline{\Cot M}\times\overline{\Cot M}\times \Cot M$$
around $G_\Delta$ which satisfy the conditions:
\begin{gather}
\label{cond1}i_\Delta\circ(i_\Delta\times \Delta_{\Cot M})  =  
i_\Delta\circ(\Delta_{\Cot M}\times i_\Delta)\\
\label{cond2}i_\Delta\circ(i_M\times\Delta_{\Cot M})=
i_\Delta\circ(\Delta_{\Cot M}\times i_M) = \Delta_{\Cot M}.
\end{gather}
We call such germs {\bf monoid structures} on $\Cot M$.
We will omit the reference to the unit morphism $\neutralmorph
\in\morph(\neutral,\Cot M)$ in the notation of a monoid 
$(\Cot M,\mu, \neutralmorph)$ 
as we have no choice for it. 

As $\Delta$ is the diagonal map, $i_\Delta$ is
a lagrangian germ around 
$$\Big\{\big((0,x),(0,x),(0,x)\big):x\in M\Big\}$$
in $\overline{\Cot M}\times\overline{\Cot M}\times\Cot M$.
Thus, the local charts
$\overline{\Cot U}\times\overline{\Cot U}\times\Cot U$ 
induced by locals charts $U$ 
of the base $M$ are enough to describe $i_\Delta$ completely . 
In the remaining of this section, we consider only such charts and
we denote by $p_1,x_1,p_2,x_2,p_3,x_3$ the local coordinates
on them.
In a local chart $V= \overline{\Cot U}\times\overline{\Cot U}\times\Cot U$,
the generating function of a monoid structure 
$i_\Delta$ is a germ of a smooth function,
$$S:B_\Delta^V = (\R^d)^*\times(\R^d)^*\times U\longrightarrow \R,$$
around the zero section which vanishes on it and such that:
$$\nabla_{p_1}S(0,0,x) = \nabla_{p_2}S(0,0,x) = x.$$
In terms of the generating function $S$, conditions 
\eqref{cond1}--\eqref{cond2} read:
\begin{gather}
\label{eq:gencond1}S\circ(S\otimes I) = S\circ(I\otimes S)\\
\label{eq:gencond2}S\circ(\neutralmorph\otimes I) =
S\circ(I\otimes\neutralmorph) = I,
\end{gather}
where by $\neutralmorph$ and $I$ stand for the generating functions 
of $(i_M,pr)$ and $(\Delta_{\Cot M},\id_M)$ respectively.
Recall from Example \ref{ex:genneutral} and Example
\ref{ex:genid} that $\neutralmorph(x) = 0$ and $I(p,x) = px$ in local charts.
We reformulate now Equations \eqref{eq:gencond1}--\eqref{eq:gencond2}
for the generating function $S$ of $(i_\Delta,\Delta)$ in a local chart.

\begin{Lemma}\label{lem:unit} 
The identity $S\circ(I\otimes e) = S\circ(e\otimes I) = I$ 
is equivalent to $S$ satisfying the following condition:
 $$S(p,0,x) = S(0,p,x) = px.$$
\end{Lemma}

\begin{proof}
We have that 
$$S\circ(e\otimes I)(p,x) = \stat{ p_1, p_2, x_1, x_2}\Big\{
S( p_1, p_2,x)+e( x_1)+I(p, x_2)- p_1 x_1- p_2 x_2\Big\}$$
The critical points are 
$ p_1 = 0$, $ p_2 = p$, $x_1 = \nabla_{p_1}S(0,p,x)$ and $
 x_2 = \nabla_{p_2}S(p,0,x).$
Thus, we get that
$S\circ(e\otimes I)(p,x) = S(0,p,x)= I(p,x) = px.$
Similarly, we obtain that $S(p,0,x) = px.$
\end{proof}

\begin{Lemma}
The identity $S\circ(S\otimes I) = S\circ(I\otimes S)$
is equivalent to the existence of  a neighborhood $N$ 
of $\{0\}^3\times U$ in $(\R^{*d})^3\times U$
where, for all $(p_1,p_2,p_3,x)\in N$, the generating function $S$ satisfies
\begin{eqnarray}\label{SGA}
S(\bar p,p_3,x) + S(p_1,p_2,\bar x) - \bar x\bar p & = &  
S(\tilde p,p_3,x) + S(p_2,p_3,\tilde x) - \tilde x\tilde p,
\end{eqnarray}
where $\bar x,\bar p,\tilde x$ and $\tilde p$ are solution
of the following implicit equations
$$\bar x = \nabla_{p_1}S(\bar p,p_3,x),\quad \tilde x = \nabla_{p_2}S(p_1,\tilde p,x)$$
$$ \bar p = \nabla_{x}S(p_1,p_2,\bar x),\quad \tilde p = \nabla_{x}S(p_2,p_3,\tilde x).$$
\end{Lemma}

\begin{proof}
We have that
$$S\circ(S\otimes I)(p_1,p_2,p_3,x) =
\stat{\bar p_1,\bar p_2,\bar x_1,\bar x_2)}
\Big\{
S(\bar p_1,\bar p_2,x)+S(p_1,p_2,\bar x_1)+I(p_3,\bar x_2)
-\bar p_1\bar x_1-\bar p_2\bar x_2
\Big\}
$$
The critical points computation yields:
$$ \bar p_1 = \nabla_xS(p_1,p_2,\bar x_1),\quad \bar x_1 = \nabla_{p_1}S(\bar p_1, \bar p_2,x)$$
$$ \bar p_2 = \nabla_x I(p_3,\bar x_2)=p_3,\quad \bar x_2 = \nabla_{p_2}S(\bar p_1, \bar p_2,x)$$
Thus, we get that
$S\circ(S\otimes I)(p_1,p_2,p_3,x) =S(\bar p,p_3,x)+S(p_1,p_2,\bar x)-\bar p\bar x,$
where $\bar p = \nabla_xS(p_1,p_2,\bar x)$ and
 $\bar x = \nabla_{p_1}S(\bar p,p_3,x).$ Similarly, one computes 
$S\circ(I\otimes S)$ directly to obtain the right-hand side
of \eqref{SGA}.
\end{proof}

The next proposition tells us that a monoid 
$(\Cot M,\mu)$ in $\Mic$ induces a 
Poisson structure on each local chart $U\subset M$
together with the local symplectic groupoid integrating it.
Let us first recall the definition of  Poisson structures
and local symplectic groupoids.

\begin{Definition}
Let $M$ be a smooth manifold. A {\bf Poisson structure} on $M$
is a Lie bracket $\{\;,\;\}$ on the algebra of smooth functions
$C^\infty(M)$ on $M$ which is a derivation in both of its arguments.
\end{Definition}
A Poisson structure may be represented by a bivector field $\alpha\in \Gamma
(\wedge^2 \Tan M)$ in the following way:
$$\{f,g\}(x) = \langle f\otimes g,\alpha\rangle.$$ In a local
chart $U$ of $M$, the bivector field $\alpha$ is represented by
a matrix $\left(\alpha^{ij}(x)\right)_{ij=1}^{\dim M}$ 
whose coefficients depend on the point $x\in U$ and which
satisfies the Jacobi identity:
$$
\alpha^{ik}\partial_k\alpha^{jl}+
\alpha^{lk}\partial_k\alpha^{ij}+
\alpha^{jk}\partial_k\alpha^{li} = 0.
$$
In local coordinates, the bracket of two functions reads:
$$\{f,g\}(x) = \alpha^{ij}(x) \partial_i f(x)\partial_j g(x).$$

\begin{Definition}
A {\bf Poisson map} $\phi:(N,\alpha_N) \rightarrow (M,\alpha_M)$
is a smooth map which preserves the Poisson bracket, i.e., such
that for $f,g\in C^\infty(M)$:
$$\phi^*\{f,g\}_M = \{\phi^*f,\phi^*g\}_N.$$
\end{Definition}
In local coordinates, the condition that $\phi:(N,\alpha_N)
\rightarrow (M,\alpha_M)$ is a Poisson map reads:
$$
\alpha^{ij}_M(\phi(x))  = \frac{\partial\phi^i(x)}{\partial x^k}\alpha_N^{kl}(x)
\frac{\partial\phi^j(x)}{\partial x^l}.
$$

The Poisson manifolds toghether with their Poisson maps form
a category, which we denote by $\Pois$.

\begin{Example}
Let $(M,\omega)$ be a symplectic manifold with
symplectic form $\omega\in\Omega(M)$. For each
function $f\in C^\infty(M)$, we associate a
Hamiltonian vector field $X_f$ by the equation 
$\iota(X_f) \omega = df$. The Poisson bracket
associated to $\omega$ is $$\{f,g\}_\omega
= \omega(X_f,X_g).$$ If $\J$ is the symplectic
matrix of $\omega$ in Darboux coordinates, 
the Poisson bivector of $\{\;,\;\}_\omega$ is
$\Ji$.
\end{Example}

\begin{Definition}
A local symplectic groupoid over a Poisson manifold $(M,\alpha)$
is a symplectic manifold $(G,\omega)$ together with an
lagrangian embedding $\epsilon:M\rightarrow G$ and two submersions
$s,t:U\rightarrow M$, defined in a neighborhood 
$U$ of $\epsilon(M)$ in $G$, such that
\begin{itemize}
\item[(1)] $s$ and $t$ are projection on $M$, i.e., $s\circ \epsilon = \id_M$
and $t\circ\epsilon = \id_M$,
\item[(2)] $s$ and $t$ are Poisson and 
anti-Poisson maps respectively, 
\item[(3)] $s$ and $t$ commute, i.e., we have that 
$\{s^*f,t^*g\}_\omega  = 0,$
for all $f,g\in C^\infty(M)$ and where $\{\;,\;\}_\omega$ is
the Poisson bracket associated to the symplectic form $\omega$.
\end{itemize}
The map $s$ is called the {\bf source} and the map $t$ is
called the {\bf target}. We write sometimes a local symplectic
groupoid over $M$ as $(s,t):G\rightrightarrows M$. 
We also say that $(s,t):G\rightrightarrows M$ integrates (in a local context)
the Poisson manifold $M$.
\end{Definition}

\begin{Remark}
Usually, the definition of symplectic groupoid $G\rightrightarrows M$ includes 
a partially defined associative product on $G$ (the product of two elements $g_1$ and $g_2$
in $G$ is defined only when $s(g_1) = t(g_2)$) whose graph is a lagrangian
submanifold of $\overline G\times\overline G\times G$. In the local case (i.e. when one requires
the source and target domains to be only a neighborhood of $\epsilon(M)$ in
$G$ and not the whole space $G$), it has been shown, in \cite{CDW1987} and in \cite{karasev1987} for instance, that it is possible
to recover the partially defined product from the data of the source and
the target maps. For this reason and
for the sake of simplicity, we prefer not to mention the partially defined product in the definition
of a local symplectic groupoid. 
\end{Remark}

\begin{Proposition}\label{prop:gpd}
Suppose that $\mu\in\morph(\Cot M^{\otimes2},\Cot M)$
is a monoid structure on $\Cot M$
whose generating function in a local chart $U$ is
$S$.
Define the bivector field $\alpha\in \Gamma(\wedge^2 \Tan U)$ by
the following matrix:
$$\alpha(x) :=\Bigg( 
\frac{\partial^2S}{\partial{p_k^1}\partial{p_l^2}}(0,0,x)-
\frac{\partial^2S}{\partial{p_l^1}\partial{p_k^2}}(0,0,x)
\Bigg)_{kl=1}^d$$
and the maps $s,t:\Cot U\rightarrow U$ by the formulas:
\[
\begin{array}{ccc}
s(p,x)      & := &   \nabla_{p_2}S(p,0,x) \\
t(p,x)      & := &   \nabla_{p_1}S(0,p,x).
\end{array}
\]
Then $\alpha\in\Gamma(\wedge^2 U)$ is a Poisson bivector on $U$, and $(s,t):\Cot M\rightrightarrows
M$ is a local symplectic groupoid integrating $\alpha$.
\end{Proposition}

\begin{proof}
We have to show that 
\begin{eqnarray}
\label{eq:Poisson}\{s^i,s^j\}_\omega(p,x) & = & \alpha^{ij}(s(p,x))\\
\label{eq:AntiPoisson}\{t^i,t^j\}_\omega(p,x) & = & -\alpha^{ij}(t(p,x))\\
\label{eq:commute}\{s^i,t^j\}_\omega(p,x) & = & 0.
\end{eqnarray}
Notice that equation \eqref{eq:Poisson} implies that $\alpha$ is a 
Poisson bivector field. Namely, Equation \eqref{eq:Poisson} means that, for any function
$f,g\in C^\infty(U)$, we have that
$$\{s^*f,s^*g\}_\omega(p,x) = s^*\{f,g\}_\alpha(p,x),$$
which yields that
$$s^*\{f,\{g,h\}_\alpha\}_\alpha = \{s^*f,\{s^*g,s^*h\}_\omega\}_\omega.$$
As $\{,\}_\omega$ fulfills the Jacobi identity and as
$s^*f(0,x) = f(x)$, we obtain that $\{,\}_\alpha$ also satisfies
the Jacobi identity.

Let us check that \eqref{eq:Poisson} holds. Derive 
Equation \eqref{SGA} two times, first with respect to $p_3$ and
then with respect to $p_2$. We obtain

$$
\frac{\partial^2S}{\partial p^1_j\partial p^2_j}(\bar p,p_3,\bar x)
\frac{d\bar p_j}{dp^2_k} =
\frac{\partial^2S}{\partial p^1_k\partial p^2_i}(p_2,p_3,\tilde x)
+\frac{\partial^2S}{\partial x^j\partial p^2_i}(p_2,p_3,\tilde x)
\frac{d\tilde x_j}{dp^2_k} 
 $$
If we set $p_1=p$, $p_2=p_3=0$, the critical points computation 
together with Lemma \ref{lem:unit} yields 
$\bar p = p$, $\tilde p=0$, $\bar x= x$, $\tilde x= s(p,x),$ and
$$
\frac{d\bar p_j}{dp^2_k} = 
\frac{\partial^2S}{\partial x^j\partial p^2_k}(p,0,x),
\quad
\frac{d\tilde x_j}{dp^2_k} =
\frac{\partial^2S}{\partial p^2_j\partial p^2_k}(p,0,x).
$$
Thus, we get
\begin{eqnarray*}
\frac{\partial^2S}{\partial p^2_i\partial p^1_j}(p,0,x)
\frac{\partial^2S}{\partial p^2_k\partial x_j}(p,0,x)
& = &
\frac{\partial^2S}{\partial p^2_i\partial p^1_k}(0,0,s(p,x))
+
\frac{\partial^2S}{\partial p^2_i\partial p^2_k}(0,0,x).
\end{eqnarray*}
Taking the difference between this last equation and
itself but with the indices $k$ and $i$ interchanged
we obtain 
$$
\frac{\partial s^k}{\partial x^j}(p,x)
\frac{\partial s^i}{\partial p^j}(p,x)
-
\frac{\partial s^i}{\partial x^j}(p,x)
\frac{\partial s^k}{\partial p^j}(p,x)
= \alpha^{ki}(s(p,x))
$$ 
which is exactly \eqref{eq:Poisson}.

The same strategy works for  \eqref{eq:AntiPoisson}
and \eqref{eq:commute}. However, for \eqref{eq:AntiPoisson}
 we have to differentiate
\eqref{SGA} with respect to $p_2$ and $p_1$ setting $p_1=p_2=0$ and
$p_3=p$. To check \eqref{eq:commute}, we have to differentiate
\eqref{SGA} with respect to $p_1$ and $p_3$ setting $p_1= p_3=0$
and $p_2=p$.
\
\end{proof}

\begin{Proposition}
Let $\mu\in\morph(\Cot M^{\otimes2},\Cot M)$ be a monoid structure
on $\Cot M$.
The Poisson bivector field $\alpha$ as well as the source map $s$ and the
target maps $t$
defined  in local charts $U\subset M$ in Proposition \ref{prop:gpd}
 glue well together on
overlapping charts and thus induce a local symplectic groupoid 
on $\Cot M$.
\end{Proposition}

\begin{proof}
Suppose two $(U_\gamma,\phi_\gamma)$ and $(U_\beta,\phi_\beta)$ are
two overlapping charts of $M$. Set $V_\gamma := \phi_\gamma(U_\gamma\cap U_\beta)$ and
$V_\beta := \phi_\beta(U_\gamma\cap U_\beta)$. We denote by $p,x$ the coordinates on
$\Cot V_\gamma$ and by $\bar p,\bar x$ the coordinates on $\Cot V_\beta$. $S_\gamma$
 (resp. $S_\beta$) is the generating function of $(i_\Delta,\Delta)$ in $U_\gamma\cap U_\beta$
expressed in the $p,x$ (resp. $\bar p,\bar x$) coordinates.
Denote by $G_{\gamma\beta}(\bar p,x) = g_{\beta\gamma}(x)\bar p$ 
the generating function
of the induced coordinate change $dg^*$ 
from $\Cot V_\beta$ to $\Cot V_\gamma$ by the
coordinate change on the base $g := g_{\beta\gamma}$ 
from $V_\gamma$ to $V_\beta$.
We know, from Lemma \ref{lem:oplus}, Lemma
\ref{lem:compform} and Proposition \ref{prop:coordchange}, that
\begin{eqnarray*}
S_\beta (\bar p_1,\bar p_2,\bar x) & = & G_{\beta\gamma}\circ S_\gamma\circ
 \left(G_{\gamma\beta}\otimes G_{\gamma\beta}\right)(\bar p_1,\bar p_2,\bar x)\\
& = & S_\gamma\circ(G_{\gamma\beta}\otimes G_{\gamma\beta})
(\bar p_1,\bar p_2,g_{\gamma\beta}(\bar x))\\
& = & 
S_\gamma(\tilde p_1,\tilde p_2,g_{\gamma\beta}(\bar x))
+g_{\beta\gamma}(\tilde x_1)\bar p_1
+g_{\beta\gamma}(\tilde x_2)\bar p_2
-\tilde p_1\tilde x_1
-\tilde p_2\tilde x_2,
\end{eqnarray*}
where $\tilde p_1,\tilde x_1,\tilde p_2$ and $\tilde x_2$
are the critical points given by the following implicit equations:
\begin{gather*}
\tilde p_1 = dg_{\beta\gamma}^*(\tilde x_1)\bar p_1,\quad
\tilde x_1 = \nabla_{p_1}S_\gamma
(\tilde p_1,\tilde p_2,g_{\gamma\beta}(\bar x))\\
\tilde p_2 = dg_{\beta\gamma}^*(\tilde x_2)\bar p_2,\quad
\tilde x_2 = \nabla_{p_2}S_\gamma
(\tilde p_1,\tilde p_2,g_{\gamma\beta}(\bar x))
\end{gather*}
Using Lemma \ref{lem:der}, we get that:
$$\nabla_{\bar p_2}S_\beta(\bar p_1,\bar p_2,\bar x)
= g_{\beta\gamma}(\tilde x_2).$$
Now, setting $\bar p_2 = 0$, we get immediately that $\tilde p_2=0$,
Lemma \ref{lem:unit} gives that $\tilde x_1 = g_{\gamma\beta}(\bar x)$
and thus 
\begin{gather*}
\tilde p_1 = dg_{\beta\gamma}^*(g_{\gamma\beta}(\bar x))\bar p_1,\quad
\tilde x_2 = \nabla_{p_2} S_\gamma(dg^*(\bar x)\bar p_1,0,g_{
\gamma\beta}(\bar x)).
\end{gather*}

Then we have  that:
\begin{eqnarray*}
s_\beta(\bar p,\bar x) & = & \nabla_{\bar p_2}S_\beta(\bar p,0,\bar x)\\
                       & = & g(\nabla_{p_2}S_\gamma((dg^*\bar p,0,g^{-1}(\bar x)))\\
                       & = & g(s_\gamma(dg^*((\bar p,\bar x))).
\end{eqnarray*}
Similarly, we get that $t_\beta(\bar p,\bar x) = g(t_\gamma(dg^*((\bar p,\bar x))).$
Thus, the $s_\gamma$'s and the $t_\gamma$'s define a global source and target on a neighborhood of $M$ in $\Cot M$.
Now, let us check the invariance of the Poisson structure $\alpha_\gamma$. 
Using Lemma \ref{lem:der}, we get that:
$$\nabla_{\bar p^1_k}S_\beta(\bar p_1,\bar p_2,\bar x)
= \nabla_{\bar p^1_k}G_{\gamma\beta}(\bar p_1,\tilde x_1)$$
and then
$$
\nabla_{\bar p^1_k}\nabla_{\bar p^2_l}S_\beta(\bar p_1,\bar p_2,\bar x)
= \nabla_{\bar p^1_k}\nabla_{x^u}G_{\gamma\beta}(\bar p_1,\tilde x_1)\frac{d \tilde x_1^u}{d \bar p_l^2}.$$
Now, if $\bar p_1 = \bar p_2 = 0$ then $\tilde p_1 = \tilde p_2 = 0$ and
$\tilde x_1 = \tilde x_2 = g^{-1}(\bar x)$ and
\begin{eqnarray*}
\left(\frac{d \tilde x_1^u}{d \bar p_l^2}\right)_{|\bar p_1 = \bar p_2 = 0}
 & = & \alpha_\gamma^{uv}(g^{-1}(\bar x))\left(\frac{d \tilde p^2_v}{d \bar p_l^2}\right)_{\bar p_1 = \bar p_2 = 0}\\
\left(\frac{d \tilde p^2_v}{d \bar p_l^2}\right)_{|\bar p_1 = \bar p_2 = 0}
& = & \frac{\partial g^l}{\partial x^v} (g^{-1}(\bar x)).
\end{eqnarray*}
Finally, we obtain the invariance of the Poisson structure, i.e.,
$$\alpha_\beta^{kl}(\bar x) =
\frac{\partial g^k}{\partial x^u} (g^{-1}(\bar x))
\alpha_\gamma^{uv}(g^{-1}(\bar x))
\frac{\partial g^l}{\partial x^v} (g^{-1}(\bar x)).$$
\end{proof}

\begin{Proposition}\label{prop:equiv1}
Let $(\Cot M,\mu_M)$ and $(\Cot N,\mu_N)$ be two monoids
 and $\alpha_M$, $\alpha_N$ their
induced Poisson structure on the base $M$ and $N$ respectively.
Suppose $T=(i_\phi,\phi)\in \morph(\Cot M,\Cot N)$ is
a monoid morphism.
Then the base map $\phi$ is a Poisson map from $(N,\alpha_N)$
to $(M,\alpha_M)$.
\end{Proposition}

\begin{proof}
Consider $U_1$ and $U_2$ two local charts of
$M$ and $N$ respectively.
Denote by $S_M$, $S_N$ and $F$ the generating 
functions, in the induced local charts, of 
$\mu_M$, $\mu_N$ and $T$ respectively.
Then we have that
\begin{eqnarray}\label{eq:genfct}
 F\circ S_M & = & S_N \circ(F\otimes F) .
\end{eqnarray}
Denote the local coordinates on $\Cot M\times\Cot M$ by
$p_1,p_2,x_1,x_2$ and the local coordinates on $\Cot N$
by $\bar p,\bar x$.
The, the left hand side of Equation \eqref{eq:genfct}
is:
$$F\circ S_M(p_1,p_2,\bar x)
= F(\tilde p,\bar x) + S_M(p_1,p_2,\tilde x) - \tilde p\tilde x
$$
where $\tilde p$ and $\tilde x$ are given by the 
following implicit equations:
$$\tilde p = \nabla_x S_M(p_1,p_2,\tilde x)\quad
\tilde x = \nabla_p F(\tilde p,\bar x).$$
By Lemma \ref{lem:der}, we obtain that:
$$
\nabla_{p_1}(F\circ S_M)(p_1,p_2,\bar x)
%\frac{\partial(F\circ S_M)}{\partial p_1} (p_1,p_2,\bar x) 
= \nabla_{p_1} S_M(p_1,p_2,\tilde x).$$
If we derive this equation again with respect to $p_2$, we get:
$$
\nabla_{p_1}\nabla_{p_2} (F\circ S_M)(0,0,\bar x)
%\frac{\partial^2 (F\circ S_M)}{\partial p_1\partial p_2} (p_1,p_2,\bar x)
= \nabla_{p_1}\nabla_{p_2} S_M(p_1,p_2,\tilde x)
+ \nabla_x S_M(p_1,p_2,\tilde x) 
\frac{d\tilde x}{d p_2}.
$$
Setting $p_1=p_2=0$, we get that $\tilde p=0$ and
$\tilde x = \phi(\bar x)$ and thus:
\begin{eqnarray*}
\nabla_{p_1}\nabla_{p_2} (F\circ S_M)(0,0,\bar x)
%\frac{\partial^2 (F\circ S_M)}{\partial p_1\partial p_2} (0,0,\bar x)
& = & \nabla_{p_1}\nabla_{p_2} S_M(0,0,\phi(\bar x))\\
& = & \alpha_M(\phi(\bar x)).
\end{eqnarray*}
Now, the right hand side of Equation \eqref{eq:genfct} yields:
$$
S_N\circ(F\otimes F)(p_1,p_2,\bar x)
= S_N(\tilde p_1,\tilde p_2,\bar x)
+ F(p_1,\tilde x_1)
+ F(p_2,\tilde x_2)
-\tilde p_1\tilde x_1
-\tilde p_2\tilde x_2,
$$
where $\tilde p_1,\tilde x_1,\tilde p_2$ and
$\tilde x_2$ are given by the following implicit
equations:
\begin{gather*}
\tilde p_1 = \nabla_xF(p_1,\tilde x_1)\quad
\tilde x_1 = \nabla_{p_1}S_N (\tilde p_1,\tilde p_2,\bar x)\\
\tilde p_2 = \nabla_xF(p_2,\tilde x_2)\quad
\tilde x_2 = \nabla_{p_2}S_N (\tilde p_1,\tilde p_2,\bar x).
\end{gather*}
Again, Lemma \ref{lem:der} gives us:
$$
\nabla_{p_1} (S_N\circ(F\otimes F))(p_1,p_2,\bar x)
%\frac{\partial S_N\circ (F\otimes F)}{\partial p_1} (p_1,p_2,\bar x) 
= \nabla_{p_1} F(p_1,\tilde x_1).
$$
Deriving another times with respect to $p_2$, we obtain:
$$
\frac{\partial^2 S_N\circ (F\otimes F)}{\partial p_i^1 \partial p_j^2}
(p_1,p_2,\bar x) = 
\frac{\partial^2 F}{\partial x^k\partial p_i^1}(p_1,\tilde x_1)
\frac{d\tilde x_1^k}{d p_j^2}.
$$
Setting $p_1 = p_2 = 0$, then $\tilde p_1 = \tilde p_2 = 0$ and
$\tilde x_1 = \tilde x_2 = \bar x$ and
$$
\frac{\partial^2 S_N\circ (F\otimes F)}{\partial p_i^1 \partial p_j^2}
(0,0,\bar x) = 
\frac{\partial \phi^i}{\partial x^k}(\bar x) 
\left(\frac{d \tilde x_1^k}{d p_j^2}\right)\Big|_{p_1=p_2=0}.
$$
Now, we have that:
$$
\frac{d \tilde x_1^k}{d p_j^2} = 
\frac{\partial^2 S_N}{\partial p_k^1\partial p_u^1}
(\tilde p_1,\tilde p_2,\bar x) 
\frac{d\tilde p_u^1}{d p_j^2}
+
\frac{\partial^2 S_N}{\partial p_k^1\partial p_u^2}
(\tilde p_1,\tilde p_2,\bar x) 
\frac{d\tilde p_u^2}{d p_j^2}.
$$
By Lemma \ref{lem:unit}, the first term of the last 
equation vanishes when $p_1=p_2=0$ and we obtain that:
$$
\left(\frac{d \tilde x_1^k}{d p_j^2}\right)\Big|_{p_1=p_2=0}
= 
\alpha_N(\bar x)^{ku}
\left(\frac{d\tilde p_u^2}{d p_j^2}\right)\Big|_{p_1=p_2=0}.
$$
In turns, we get:
$$
\frac{d\tilde p_u^2}{d p_j^2}
=
\frac{\partial^2 F}{\partial x^u\partial p_j^2}
(p_2,\tilde x_2) + 
\frac{\partial^2 F}{\partial x^u\partial x^v}
(p_2,\tilde x_2)
\frac{d\tilde x_2^v}{d p_j^2},
$$
which yields:
$$
\left(\frac{d\tilde p_u^2}{d p_j^2}\right)\Big|_{p_1=p_2=0}
= \frac{\partial \phi^j}{\partial x^u}(\bar x).
$$
Finally, we obtain:
$$
\frac{\partial^2 \big(S_N\circ (F\otimes F)\big)}{\partial p_i^1 \partial p_j^2}
(0,0,\bar x) =
\frac{\partial \phi^i}{\partial x^k}(\bar x)
\alpha_N^{ku}(\bar x)
\frac{\partial \phi^j}{\partial x^u}(\bar x).
$$
As $F\circ S_M = S_N\circ(F \otimes F)$, we conclude that:
$$\alpha_M^{ij}(\phi(\bar x)) = 
\frac{\partial \phi^i}{\partial x^k}(\bar x)
\alpha_N^{ku}(\bar x)
\frac{\partial \phi^j}{\partial x^u}(\bar x),
$$
which means that $\phi$ is a Poisson map form $(N,\alpha_N)$ to
$(M,\alpha_M)$.
\end{proof}

We may now define the Poisson functor
$$\Deq:\Mon\Mic\longrightarrow \Pois,$$
by assigning to each monoid $(\Cot M,\mu_M)$
the Poisson manifold $(M,\alpha_M)$ as in
Proposition \ref{prop:gpd} and by assigning to
each monoid morphism
$$T=(i_\phi,\phi):(\Cot M,\mu_M) \longrightarrow (\Cot N, \mu_N)$$
the map $\phi: (N,\alpha_N)\rightarrow (M,\alpha_M)$. Proposition
\ref{prop:equiv1} garantees that $\phi$ is a Poisson map. The functoriality
of $\Deq$ follows directly from the properties of map composition.

\begin{Definition}
Let $(s_M,t_M):G_M\rightrightarrows M$ and
$(s_N,t_N):G_N\rightrightarrows N$ be two local
symplectic groupoids. An isomorphism between
local symplectic groupoids is a germ of 
symplectomorphisms $\psi:G_M\rightarrow G_N$
around $M$ which sends $M$ to $N$ and such
that:
\begin{eqnarray}
\label{def:source}s_N \circ \Psi & = & \Psi\circ s_M\\
\label{def:target}t_N \circ \Psi & = & \Psi\circ t_M.
\end{eqnarray}

\end{Definition}
\begin{Proposition}
Under the same assumption as in Proposition \ref{prop:equiv1},
suppose further that $T=(i_\phi,\phi)$ is invertible. Then the
transverse lagrangian germ $i_\phi$ comes from 
the graph of a germ $\Psi$ of symplectomorphisms around $Z_M$, 
and preserving the bases. Moreover, $\Psi$ is an isomorphism  
between the induced local symplectic groupoids.
\end{Proposition}

\begin{proof}
Let $\Psi$ be the germ of symplectomorphism induced by $T$
as in Proposition \ref{prop:germinv}. Denote by $S_M$, $S_N$ and $F$ 
the generating function of $\mu_M$, $\mu_N$ and $T$ in a local
chart and denote by $p_1,x_1,p_2,x_2$ the local coordinates
on $\Cot M\times\Cot M$. By definition, we have that:
$$\Psi\Big(p_1,\nabla_pF(p_1,x_2)\Big)=
\Big(\nabla_x F(p_1,x_2),x_2\Big).$$
Verifying Equation \eqref{def:source} is then equivalent
to verifying that:
$$s_N\big(\nabla_xF(p_1,x_2),x_2\big)
= \Psi\Big(0,s_M\big(p_1,\nabla_p F(p_1,x_2)\big)\Big).
$$
This is equivalent to see that:
\begin{gather}\label{eq:toverify}
\nabla_{p_2} S_N\Big(\nabla_xF(p_1,x_2),0,x_2\Big) = 
\phi^{-1}\Big(\nabla_{p_2}S_M\big(p_1,0,\nabla_p F(p_1,x_2)\big)\Big).
\end{gather}
Now, Lemma \ref{lem:der} gives us that:
$$
\nabla_{p_2}(F\circ S_M)(p_1,p_2,\bar x)
= \nabla_{p_2}S_M(p_1,p_2,\tilde x)
$$
where $\tilde x$ is defined by the implicit equations
for $\tilde p$ and $\tilde x$:
$$\tilde p = \nabla_x S_M(p_1,p_2,\tilde x)
\quad
\tilde x = \nabla_p F(\tilde p,\bar x).$$
Setting $p_2=0$, we get, by Lemma \ref{lem:unit}, that $\tilde p = p_1$
and $\tilde x = \nabla_p F(p_1,\bar x)$. Thus,
$$\nabla_{p_2}(F\circ S_M) 
(p_1,p_2,\bar x) = \nabla_{p_2}S_M(p_1,0,\nabla_p
F(p_1,\bar x)).$$
On the other hand, Lemma \ref{lem:der} tells us that:
$$\nabla_{p_2}\big(S_N\circ(F\otimes F)\big)
(p_1,p_2,\bar x)
= \nabla_{p_2} F(p_2,\tilde x_2),$$
where $\tilde x_2$ comes from the solution of the
implicit system for $\tilde p_1,\tilde x_1,\tilde p_2$
and $\tilde x_2$: 
\begin{gather*}
\tilde p_1 = \nabla_xF(p_1,\tilde x_1)\quad
\tilde x_1 = \nabla_{p_1}S_N (\tilde p_1,\tilde p_2,\bar x)\\
\tilde p_2 = \nabla_xF(p_2,\tilde x_2)\quad
\tilde x_2 = \nabla_{p_2}S_N (\tilde p_1,\tilde p_2,\bar x).
\end{gather*}

Setting $p_2 = 0$, we get that $\tilde p_2 = 0$, 
$\tilde x_1 = \bar x$, $\tilde p_1 = \nabla_x F(p_1,\bar x)$ and thus:
$$\tilde x_2 = \nabla_{p_2}S_N\big(\nabla_xF(p_1,\bar x),0,\bar x\big).$$

Thus we get that:
$$\nabla_{p_2}\big(S_N\circ(F\otimes F)\big)
(p_1,0,\bar x) =
\phi(\tilde x_2) = \phi\Big(
\nabla_{p_2}S_N\big(\nabla_xF(p_1,\bar x),0,\bar x\big)
\Big).
$$
Finally, the fact that $F\circ S_M = S_N\circ(F\otimes F)$
implies \eqref{eq:toverify}.
\end{proof}

Let us summarize the content of this section in the following
theorem.

\begin{Theorem}\label{thm:morph}
In $\Mic$, a monoid $(\Cot M,\mu)$
induces a Poisson structure on the base $M$ together with 
a local symplectic groupoid on $(s,t):\Cot M\rightrightarrows M$
integrating it. Isomorphisms of monoids induce
Poisson diffeomorphisms between the induced Poisson structures
and local symplectic groupoid isomorphisms between the induced
local symplectic groupoids. 
\end{Theorem}

\section{Examples}\label{sec:ex}

In this Section, we describe explicitly some
examples of monoid structures on cotangent
bundles. We provide formulas for their induced
Poisson structures and local symplectic groupoids.

\subsection{Symplectic manifolds}
Let $(\R^{2n},\J)$ be the standard symplectic
manifold. We consider its cotangent bundle $\Cot \R^{2n}$
as an object in the cotangent microbundle category. We construct
a monoid structure $\mu = (i_S,\Delta)\in \morph((\Cot \R^{2n})^{\otimes
2},\Cot \R^{2n})$ on it thanks to the symplectic matrix $J$.
The transverse lagrangian germ 
$$i_S:B_\Delta \hookrightarrow \overline{\Cot \R^{2n}}\times
\overline{\Cot \R^{2n}}\times \Cot \R^{2n}$$
is given by the following generating function 
\begin{eqnarray}\label{genfct:sympl}
S(p_1,p_2,x) & = & (p_1+p_2)x + \frac12p_1^T\Ji p_2
\end{eqnarray}
where $\Ji$ is the inverse of $J$.
A straightforward computation yields:
\begin{eqnarray*}
M & = & \big(S\circ(S\otimes I)-S\circ(I\otimes S)\big)(p_1,p_2,p_3,x)\\
& = & \frac12\big(p_1^T\Ji p_2 + (p_1+p_2)^T\Ji p_3\big)
      - \frac12\big(p_1^T\Ji(p_2+p_3) + p_2^T\Ji p_3\big)\\
  & = & 0,
\end{eqnarray*}
which means that $(i_S,\Delta)$ is a monoid structure on
$\Cot \R^{2n}$. Note that the induced Poisson structure, 
$$
\left(\frac{\partial^2 S}{\partial p_i^1\partial p_j^2}(0,0,x)-
\frac{\partial^2 S}{\partial p_j^1\partial p_i^2}(0,0,x)\right)
 = (\Ji)^{ij},
$$ 
is the inverse $\Ji$ of the original symplectic form $J$.
The induced source and target $(s,t):\Cot \R^{2n}\rightrightarrows \R^{2n}$
are given by the formulas:
$$
s(p,x) = x + \frac12 \J p\quad\textrm{and}\quad
t(p,x) = x - \frac12 \J p.
$$

\begin{figure}[h!]
\labellist
\small\hair 2pt
\pinlabel $x_1$ at -4 0 
\pinlabel $x_2$ at 77 60 
\pinlabel $x_3$ at 141 0 
\pinlabel $\Ji(p_1)$ at 30 38 
\pinlabel $\Ji(p_2)$ at 115 38
\pinlabel ${\Ji (p_1+p_2)}$ at 70 8 
\endlabellist
\centering
\includegraphics[scale=1]{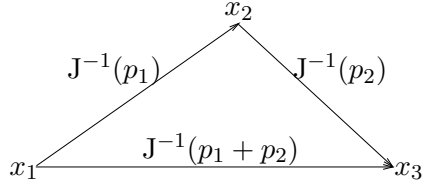}
\caption{The triangle $T(p_1,p_2,x)$}
\label{fig:triangle}
\end{figure}

There is a nice geometric interpretation of both the generating
function \eqref{genfct:sympl} and the associativity
equation reminiscent of \cite{dROA2004}
and \cite{weinstein1994}. Let us consider $\R^2$ with
its standard symplectic form $\J$ for simplicity. To each
point $(p_1,p_2,x)\in B_\Delta$, we may associate a triangle
$T(p_1,p_2,x)$ in $\R^2$ in the following way. Consider the Hamilton
flows on $\R^2$, $\Psi_1^t$ and $\Psi_2^t$, of the linear Hamiltonians
$l_{p_1}(x) = p_1x$ and $l_{p_2}(x) = p_2x$ respectively. 
The three vertices of the triangle are given by
$x_1 = x$, $x_2 = \Psi_1^t|_{t=1}(x_1)$ and
$x_3 = \Psi_2^t|_{t=1}(x_2)$. The edge joining $x_1$ to
$x_2$ is the trajectory of $x_1$ under $\Psi_1^t$ and the
edge joining $x_2$ to $x_3$ is the trajectory of 
$x_2$ under $\Psi_2^t$. One can verify that the 
Hamilton flow of the Hamiltonian $l_{p_1+p_2}$ carries
$x_1$ to $x_3$ along the third edge of the triangle. An alternative
description of $T(p_1,p_2,x)$ is the triangle
with vertex $x$ and defined by the two vectors
$\Ji p_1$ and $\Ji p_2$ as in Figure \ref{fig:triangle}.
 
\begin{figure}[h!]
\labellist
\small\hair 2pt
\pinlabel $x$ at -4 -2 
\pinlabel $x$ at  196 -2 

\pinlabel $\Ji(p_1)$ at 5 33
\pinlabel $\Ji(p_2)$ at 72 62
\pinlabel $\Ji(p_3)$ at 140 33
\pinlabel ${\Ji (p_1+p_2+p_3)}$ at 65 -5

\pinlabel $\Ji(p_1)$ at 205 33
\pinlabel $\Ji(p_2)$ at 272 62
\pinlabel $\Ji(p_3)$ at 340 33
\pinlabel ${\Ji (p_1+p_2+p_3)}$ at 265 -5

\pinlabel ${\frac12 p_1^T\Ji p_2}$ at 57 45
\pinlabel ${\frac12 (p_1+p_2)^T\Ji p_3}$ at  85 16
\pinlabel ${\frac12 p_2^T\Ji p_3}$ at 290 45 
\pinlabel ${\frac12 p_1^T\Ji (p_2+p_3)}$ at 257 16

\endlabellist
\centering
\includegraphics[scale=1]{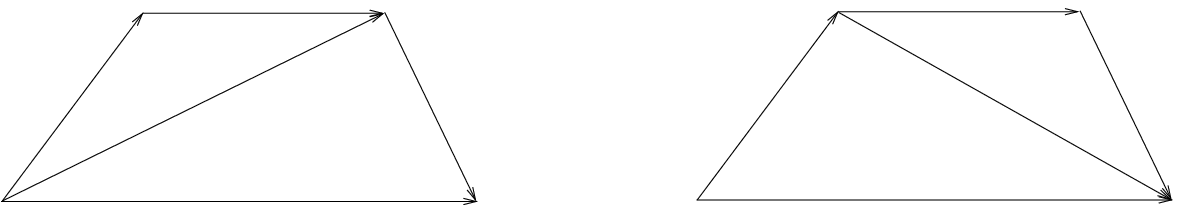}
\caption{The associativity equation in terms of areas.}
\label{fig:assos}
\end{figure}

The area $A(p_1,p_2,x)$ of $T(p_1,p_2,x)$ is given by the
formula:
$$\frac12\det(\Ji p_1,\Ji p_2) = \frac12 p_1^T\Ji p_2.$$
The generating function $S$ may then be written as:
$$S(p_1,p_2,x) = (p_1+p_2)x + \operatorname{Area}\Big(T(p_1,p_2,x)\Big).$$
The associativity equation may be interpreted
as an equality between areas as shown in Figure \ref{fig:assos}.
 
\subsection{Lie algebras}
We consider the cotangent bundle $\Cot \R^d$ and look
for monoid structures $$i_\Delta:B_\Delta
\rightarrow \overline{\Cot \R^d}\times\overline{\Cot \R^d}
\times \Cot \R^d$$
whose generating function $S_\Delta:B_\Delta\rightarrow \R$
is linear in $x$:
$$S(p_1,p_2,x) = \langle x, A(p_1,p_2) \rangle.$$
Note that $S$ being a germ of functions around 
the zero section and which vanishes on it implies
that $$A:(\R^d)^*\times (\R^d)^*\longrightarrow (\R^d)^*$$
must be a germ of a map around $(0,0)$ and such that
$A(0,0) = 0$. The equation $$S\circ (\neutralmorph\otimes I) = I =
S\circ (I\otimes\neutralmorph)$$ implies by Lemma \ref{lem:unit}
that $$A(p,0) = A(0,p) = p.$$ A straightforward computation
tells us that the associativity equation,
\begin{eqnarray*}
M & = & \big(S\circ(S\otimes I)-S\circ(I\otimes S)\big)(p_1,p_2,p_3,x)\\
& = &  \langle x, A(p_1,A(p_2,p_3)- A(A(p_1,p_2),p_3) \rangle\\
& = & 0,
\end{eqnarray*}
is equivalent to the associativity of the map $A$.
The induced Poisson structure is given by:
$$\alpha^{ij}(x) = 
 \left(
\frac{\partial^2 A_k}{\partial p_i^1\partial p_j^2}(0,0) -
\frac{\partial^2 A_k}{\partial p_j^1\partial p_i^2}(0,0)
\right) x^k,$$
which is a linear Poisson structure on $\R^d$. This implies,
in particular, that
$$C_k^{ij} = 
 \left(
\frac{\partial^2 A_k}{\partial p_i^1\partial p_j^2}(0,0) -
\frac{\partial^2 A_k}{\partial p_j^1\partial p_i^2}(0,0)
\right) 
$$
are the structure constants of a Lie algebra structure on $\R^d$.
We denote this Lie algebra by $\mathcal G$.
The source and target are given by the formulas:
$$
s(p,x) = \langle x,\nabla_{p_2}A(p,0)\rangle \quad\textrm{and}\quad
t(p,x) = \langle x,\nabla_{p_1}A(0,p)\rangle.
$$

Conversely, if we start from a Lie algebra $(\mathcal G,[\;,\;])$,
consider the Baker-Campbell-Hausdorff map:
$$BCH: \mathcal O\times \mathcal O \longrightarrow \mathcal O$$
defined in a neighborhood $\mathcal O$ of $0$ in $\mathcal G$ by
$$BCH(p_1,p_2) = \log(\exp(p_1)\exp(p_2)),$$
where  $\exp$ is the usual diffeormorphism one can construct between sufficiently small 
neighborhoods of $0$ in $\mathcal G$ and neighborhoods of the unit element 
$e$ in the corresponding Lie group $G$ and where $\log$ stands for its inverse.
%$$BCH(p_1,p_2) = p_1+p_2 + \frac12[p_1,p_2]+
%\frac1{12}\big([p_1,[p_1,p_2]]+ [p_2,[p_2,p_1]]\big)+ \cdots
%$$
The $BCH$ map provides a generating function $S:
\mathcal G\oplus \mathcal G\oplus \mathcal G^*\rightarrow \R$
of the above form, i.e., 
\begin{eqnarray}\label{BCH}
S(p_1,p_2,x) & = & \langle x, BCH(p_1,p_2)\rangle.
\end{eqnarray}
This gives a monoid structure on $\Cot{} \mathcal G^*$.
The induced Poisson structure on $\mathcal G^*$ is the
Kirillov-Kostant Poisson structure associated to the
Lie bracket of $\mathcal G$.

\subsection{Kontsevich's star-product}
Consider an open subset $U$ of $\R^d$ endowed with an analytic
 Poisson structure $\alpha$. We will describe here a monoid structure
 on $\Cot \R^d$
which induces the Poisson structure $\alpha$ and encompasses
the two previous examples, i.e., when $\alpha$ comes
from a symplectic structure $\J$ and when $\alpha$ comes
from a Lie algebra. Consider the following formal power
series in $\epsilon$:
\begin{eqnarray}\label{unigenfct}
S(\alpha)(p_1,p_2,x) & = & (p_1+p_2)x + \sum_{n=1}^\infty
\frac{\epsilon^n}{n!} \sum_{\Gamma \in T_{n,2}}
W_\Gamma \hat B_\Gamma(\alpha)(p_1,p_2,x),
\end{eqnarray}
where $T_{n,2}$ are the Kontsevich trees of type $(n,2)$
and $W_\Gamma$ their associated Kontsevich weights. 
The $\hat B_\Gamma$ are the symbols of the Kontsevich
bidifferential operators $B_\Gamma$, defined by the formula:
$$B_\Gamma(e^{p_1x},e^{p_2x}) = \hat B_\Gamma(p_1,p_2,x)e^{(p_1+p_2)x},$$
where $p_1,p_2\in(\R^d)^*$ and $x\in\R^d$. We refer the reader
to \cite{CDF2005} and \cite{kontsevich1997} for more details concerning the 
construction of formula \ref{unigenfct}.
In \cite{dherin2006}, it has been shown that 
\eqref{unigenfct} converges in a neighborhood of $Z_\Delta$
in $B_\Delta$ for $\epsilon \in(0,1)$ for analytic Poisson
structures and thus produces
a transverse lagrangian germ 
$$i_{S(\alpha)}:B_\Delta \hookrightarrow
\overline{\Cot U}\times
\overline{\Cot U}\times
\Cot U.
$$
In \cite{CDF2005}, it has been shown, although not in the
same language, that $S(\alpha)$ satisfies both:
\begin{gather*}
S\circ(S\otimes I) = S\circ(I\otimes S)\\
S\circ(\neutralmorph\otimes I) =
S\circ(I\otimes\neutralmorph) = I.
\end{gather*}
Thus, the associated germ $i_{S(\alpha)}$ produces
a monoid structure on $\Cot U$. The induced Poisson
structure is the original one times $\epsilon$, i.e.,
$\epsilon\alpha$. When $\alpha$ is the inverse of
a symplectic structure $\J$, one verifies that
we get back \eqref{BCH}. When $\alpha$ comes from
a Lie algebra, one gets back \eqref{genfct:sympl}.
The generating function \eqref{unigenfct}, may be
considered as the semi-classical part of Kontsevich's
star-product as constructed in \cite{kontsevich1997}
as it involved only the tree-level part of the star-product.
Namely, Kontsevich star-product may be put into the
following form (see \cite{CDF2005}). For $f,g\in C^\infty(\R^d)$, 
$$f\star g(x) = 
\exp\Big( \frac 1\epsilon \sum_{l=0}^\infty \epsilon^l
K_l(\epsilon \frac{\partial}{\partial y},
\epsilon \frac{\partial}{\partial z},x)\Big) f(y)g(z)\big|_{y=z=x},
$$
where $K_l = \sum_{\Gamma\in G_l} W_\Gamma\hat B_\Gamma$
is a sum over the Kontsevich graphs with two
ground vertices and %of type $(n,2)$ 
%***WHERE is $n$?***
with
$l$ loops.  
$K_0$ is exactly the generating function
in \eqref{unigenfct}.

\end{document}